# Room Temperature Electrochemical Synthesis of Hg-1212 Superconducting Thin Films

D. D. Shivagan, P. M. Shirage, L. A. Ekal and S. H. Pawar*
School of Energy Studies,
Department of Physics,
Shivaji University, Kolhapur- 416 004 (INDIA).
*E-mail: pawar_s_h@yahoo.com
shpawar_phy@unishivaji.ac.in

**Abstract**

In the present investigation, the novel two-step electrochemical process of room temperature synthesis of Hg-1212 superconducting films has been developed and reported first time. Electrochemical parameters were optimized by studying cyclic voltammetry (CV), linear sweep voltammetry (LSV) and chronoamperometry (CA) for the deposition of Hg-Ba-Ca-Cu alloy at room temperature. Current time transient showed progressive growth with hemispheriodal granules, which were then revealed by scanning electron microscopy (SEM). Stoichiometric electrocrystallization to get $Hg_1Ba_2Ca_1Cu_2O_{6+x}$ (Hg-1212) was completed by electrochemically intercalating oxygen species into Hg-Ba-Ca-Cu alloy at room temperature. The oxygen content in the samples was varied by varying the electrochemical oxidation period and the changes in the crystal structure, microstructure, and superconducting transition temperature ($T_c$) and critical current density ($J_c$) were recorded. The films oxidized for 28 min showed $T_c$ = 104.7 K with $J_c$ = 1.437 x $10^3$ A/cm$^2$. The dependence of superconducting parameters on oxygen content is correlated with structure property relations and reported in this paper.

# 1. Introduction

Hg-based superconductors are attracting much attention among other cuprates due to its high superconducting transition temperature, $T_c$, of 134 K and 164 K measured at ambient and high pressure of 30 GPa, respectively [1,2]. The first member $HgBa_2CuO_{4+\delta}$, of this $HgBa_2Ca_{n-1}Cu_nO_{2n+2+\delta}$ [Hg-12(n-1)n] series, was invented in 1993 by Putilin *et al.* [3] exhibiting superconductivity at 94 K. Soon after, Schilling *et al.* [1] observed superconductivity at 133.5 K in the Hg-Ba-Ca-CuO system and reported the formation of Hg-1212 and Hg-1223 phases. The highest values of $T_c$'s, which were determined as the onset temperature of Meissner signal, were 97 K [4], 127 K [5], 135 K [6], and 123 K [7] for Hg-1201, Hg-1212, Hg-1223 and Hg-1234 phases, respectively. Particular importance of the series is very high $T_c$ values than other cuprate superconductors.

Of these, Hg-1212 with $T_c = 127$ K is relatively stable phase bearing the highest $T_c$ than obtained for 1212 phase of the other cuprates and hence extensively studied for the fundamental studies, particularly of irreversibility line (IL) and anisotropy problems [8-9]. For the fundamental studies of basic intrinsic properties one must be in a position to synthesize these novel materials in thin film form. Further, to achieve desired device performance, technologically reproducible high quality single-phase films are required. Recently, the Hg-1212 films with $T_c = 120$ K exhibiting a critical current density, $J_c$, of about $10^6$ A/cm$^2$ at 77 K [10] have successfully synthesized. The first ever-superconducting quantum interference device (SQUID) operating above 110 K was made from these Hg-1212 films [11]. This offers the possibility of electronic and microelectronic applications of these superconducting films above 110 K.

For the synthesis of superconducting thin films, the range of processes such as chemical vapor deposition, laser ablation, molecular beam epitaxy, magnetron sputtering, physical vapor deposition, sol gel, spray pyrolysis and electrodeposition have been used. In addition to the experimental complications, the following two problems make synthesis of mercurocuprates a difficult task. First, notably, the high volatile nature of Hg and HgO oxides at relatively low temperatures of 360-500 $^{o}$C, depending on the process technique. Besides the severe problem of reduction in Hg stoichiometry, the evolution of this poisonous toxic mercury oxide vapour are dangerous for living being and creates environmental problem. Second, the formation of stoichiometric $Ba_2Ca_{n-1}Cu_nO_x$ precursor requires very high annealing temperatures of about 680- 880 $^{o}$C for longer times depending upon the process [12]. Further these precursors are very sensitive to air and humidity [13].

To avoid this, most of the researchers have adopted the two-step process as given below [13]:

i) Precursor pellets/ films of $Ba_2Ca_{n-1}Cu_nO_x$ were synthesized by annealing at high temperature.

ii) Mercury was deposited or precursor was subjected in a controlled Hg-atmosphere.

When such multi-steps are introduced in the process it becomes difficult to maintain the process condition to get reproducible and good quality films.

To avoid this, one must choose a reliable process such as electrodeposition: a soft solution process. In this technique, as the reaction takes place on an atomic level, the Gibbs free energy required in the thermodynamic process for the formation of crystalline materials is being provided by the reaction kinetics. Hence avoiding the powder calcining approach it gives the stoichiometric good quality films at room temperature and in less time.

In the present investigation, an attempt has been made to synthesize $HgBa_2CaCu_2O_{6+\delta}$ superconducting films by two step electrodeposition technique, at room temperature. Electrochemical parameters of individual constituents were studied and the processing of the alloy was optimized by cyclic voltammetry (CV), linear sweep voltammetry (LSV), and chronoamperometry techniques. Pulsed potential technique was employed to achieve the control over the desired atomic ratio and morphology of the deposited alloy. Further, a novel electrochemical oxidation technique, which makes it possible to oxidize the films at room temperature, was used to oxidize the alloy films with desired stoichiometry to become superconducting. The films were characterized by XRD, SEM and EDAX techniques for structural, morphological and compositional studies, respectively. The superconducting parameters such as $T_c$ and $J_c$ were measured and results are reported in this paper.

## 2   Electrochemistry of Alloy Deposition:

Electrodeposition is a process for depositing the metals, alloys and oxides on a conducting substrate from a bath containing the ions of interest. In recent years, an increasing interest in electronic industry using electrodeposition for microfabrication purposes and in the surface treatment industry confronted with the need for the development of new types of functional coatings that are environmentally safe.

In a bath containing metal ions, when a sufficient negative potential greater than the reduction of metal ions is applied, the metal ions get deposited onto the cathode surface. The equilibrium reduction potential, $E_m$, of the metal electrode (cathode) in a given solution is given by the familiar Nernst equation,

$$E_m = E_m^o + (RT/mF) \ln [(a_M^{m+})/a_M] \qquad (1)$$

where,

$E_m$ - is standard potential to form M, 'R' – is gas constant, 'T'- is absolute temperature, 'm'- is the number of electrons required for the reduction, 'F'- is the faradays constant, and '$a_M^{m+}$' and '$a_M$' are activities (concentrations) of metal ions in the electrolyte and in the deposit respectively.

Electrodeposition of 'M' can occur at potentials more negative than the equilibrium potential; this difference in potential is the overpotential (or over voltage), $\eta$. The rate and amount of deposited metal can be monitored by adjusting the metal ion concentration in the bath or by applying sufficient overpotentials, which is governed by the following equation,

$$d[M^o]/dt = k_m [M^{n+}] \qquad (2)$$

where, $k_m$ is the potential-dependent rate constant and $[M^{n+}]$ is the solution concentration of the metal ions.

It is easier to deposit the single metal if one takes an account of the interactions of the solute ions $M^{m+}$ with solvent, or with complex formation. The interfacial activity of $M^{m+}$, which depends on ionic strength, must be carefully controlled.

In case of alloy or compound deposition, it is required to deal with two or more metals having different equilibrium deposition potentials. Hence, for the alloy deposition, equilibrium potentials, activities of the ions in the solution and the stability of the resultant alloy are to be understood crucially [16]. For example, for compound $M_mN_n$ deposition, component M involves 'm' electrons and that of component N, 'n' electrons. The equilibrium potentials of M and N can be determined by Nernst equation and are different.

However, simultaneous deposition of two different kinds of ions at the cathode can be possible by keeping following condition.

$$E_m + \eta^m = E_n + \eta^n \quad (3)$$

The activities of the metals M and N in the compound or alloy are determined by their concentrations, and by the thermodynamical stability of the deposit. The reversible potential of a metal M alloyed with the component N should be more positive than that of pure metal. This is because of the free energy of formation of the alloy or compound, $\Delta G$, with a shift of potential [14],

$$E = -\Delta G / rmF \quad (4)$$

where, 'm' is the constituents valence, 'r' is mole fraction, 'F' is faradays constant and '$\Delta G$' is the free energy.

The shift of potential is a constant value for the formation of a compound, but varies with composition in the case of alloy formation. This often makes it very difficult to control atomic ratio in deposit. The significant variation of electrochemical potential can be expected [15].

Further, different kinetic approaches based on kinetic concept and considering the nature of coupling phenomenon between co-depositing species, following types of behaviours are discussed recently [17]. In 'non-interactive co-deposition', the partial current densities of the co-depositing metals are largely independent of each other. The co-deposition of copper and nickel under such conditions behave in this way [18]. Often in alloy deposition, the partial current densities at the metal electrolyte interface are coupled i.e. the rate of charge transfer of a given species 'A' depends on that of the co-depositing species 'B' called as 'charge transfer coupled co-deposition'. The co-deposition of 'B' may lead to decrease, 'inhibited co-deposition', or to an increase, 'catalyzed co-deposition' of the deposition rate of 'A' and vice versa. On the other hand, the 'induced' co-deposition can be explained by a catalytic effect of the codepositing metal. Finally, the co-deposition reactions are coupled through mass transport process, 'mass transport coupled co-deposition'. For example, the reduction of a complexed species may release legends at the cathode surface that affects the complexing equilibrium and hence the reduction rate of codepositing metals.

In the complexing ion deposition technique, the codepositing species are simultaneously reduced after the complex reduction potential. Pawar *et al*. have obtained Dy-Ba-CuO, Sm-Ba-CuO, YBCO and BSCCO [19-24] based superconducting thin films by employing the various complexing baths in aqueous solution. Further, the technique was modified by using the non-aqueous solvents such as DMSO [22-24]; and simply by adjusting the individual bath concentrations and complex reduction over-potentials based on the empirical facts, good quality films were achieved.

## 3   Experimental Procedure

Electrolytic bath was prepared by dissolving mercuric chloride and reagent grade nitrates of barium, calcium and copper in dimethyl sulphoxide (DMSO). The conventional three-electrode system consisting of saturated calomel electrode (SCE) as a reference electrode, graphite as a counter electrode and mirror polished silver foil as working electrode was used. Initially, the individual baths with 50 mM concentrations of Hg, Ba, Ca and Cu were prepared and cyclic voltammetry of individual baths and DMSO was studied using VersaStat-II in order to understand the possible reactions at electrode. Then linear sweep voltammograms (LSV) of individual baths and combined bath were recorded to estimate the deposition potentials. After LSV study, a complexing bath was selected to be 45 mM $HgCl_2$, 80 mM $Ba(NO_3)_2$, 65 mM $Ca(NO_3)_2 \cdot 2H_2O$ and 50 mM $Cu(NO_3)_2 \cdot 3H_2O$ to get HgBaCaCu alloyed films with atomic ratio 1:2:1:2. Now the potentiostatic square wave pulse generator (model 1130) with variable frequency from 1 Hz to 10 MHz and duty cycle from 1 to 100 % was used. The square wave pulse with 25 Hz frequency and 50 % duty cycle was optimized and employed in the present pulse deposition process. From cathodic polarization curve and variation in current density with time for different potentials, the deposition potential was optimized to be –1.7 V vs. SCE. The films were deposited under potentiostatic conditions for different lengths of time and thickness was measured by gravimetric weight difference-density method.

The as-deposited alloyed films were then electrochemically oxidized from alkaline 1 N KOH solution at a potential of +0.7 V vs. SCE at room temperature. The EG & G scanning potentiostat (model 362) and VersaStat-II was used for the electrochemical oxidation. X-ray diffraction patterns of as-deposited and electrochemically oxidized films were recorded using microcomputer controlled Phillips-3710 diffractometer with $CuK_\alpha$ radiations. The microstructural measurements were done using scanning electron microscope (CAMECA model-30) attached with EDAX. The electrical resistivity and critical current density $J_c$ were measured using standard four probe resistivity technique, where contacts were made by air-drying silver paste. The samples were cooled in 10 K He close cycle refrigeration system. A constant current was passed through the current contacts and the voltage developed was measured with Keithley multimeters and temperature was measured with the help of silicon diode sensor.

# 4 Experimental Results

## 4.1 Electrochemistry of Deposition of HgBaCaCu (1:2:1:2) Alloyed Films

The interface between the electrode and an electrolyte is the heart of electrochemistry. It is the place where the charge transfer takes place and gradients in electrical and chemical potentials constitute the driving force for the electrochemical reactions and hence the alloy formation. The classic route to the study of electrochemical reactions rests on current and voltage measurements, i.e. cyclic voltammetry (CV) and linear sweep voltammetry (LSV). The CV and LSV studies give the electrochemical reduction/oxidation potentials of the electrodepositing constituents based on the thermodynamic considerations and governed by the Nernst equations of equilibrium potentials. Hence, this approach was followed in determining the deposition potentials and bath concentrations.

### *4.1.1 Cyclic Voltammetry (CV)*

The interpretation of cyclic voltammogram for any electrodeposition system is often not straightforward owing to the combined influence of a number of processes involving charge transfer, adsorption, electrodeposition and coupled chemical reactions. CV has still proved to be a very popular and useful probe, particularly when one is studying a new system.

To understand the individual reduction potentials and whether the constituting metal species do undergo any oxidation/reduction or multi-step reactions, the cyclic voltammetry for individual metal bath in dimethyl sulphoxide (DMSO) was studied. The individual baths were prepared with 50 mM of $HgCl_2$, and nitrates of Ba, Ca, and Cu. The Ag foil was used as substrate and CV was recorded for 0 to -3.0 V vs. SCE with the scan rate of 20 mV/sec. This potential range was selected by considering the individual standard reduction potentials.

Figure 1 shows the CVs recorded onto Ag substrate for individual baths. The CV recorded only for non-aqueous solvent (DMSO), showed no reduction or oxidation reactions in the applied cathodic potential range. This might be due to its high dielectric constant [25] and is sufficiently resistant to both oxidation and reduction, providing a fairly wide working potential range.

In the CV recorded for $HgCl_2$ bath, it is observed that the magnitude of current is fairly larger than that for DMSO representing the ionic current, but the sudden increase in current due to liberation of ions is not evidenced. Hence as it was expected to get $Hg^{2+}$ reduction, it is bit difficult to determine cathodic potential where the Hg reduction takes place by the following reaction,

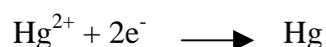

$$Hg^{2+} + 2e^- \longrightarrow Hg$$

Or whether Hg does reduced or not? This was resolved by the early studies of electrochemistry of mercury.

The mercury in chloride solution was observed to be reduced at positive potential, particularly when calomel electrode was used in the electrochemical cell [22]. This implies that some species other than $Hg_2^{2+}$ were involved in the functioning of the reduction process.

Hill and Ives [26] in 1951 concluded that the main ionic entities in equilibrium with the calomel electrode are chloride, chloromercurous ions ($Hg_2Cl^+$) and chloromercuric ions ($HgCl^+$). Some of these ions come into equilibrium and as a result so-called chloromercury is formed which consists of monolayer of chlorine atoms covalently bound to the mercury surface. Further, Erdey-Gruz and Volmer [27] have nucleated stable layer of liquid mercury on metal surface by applying certain overpotential of –0.34 V vs. SHE for platinum and –0.21 V vs. SHE for carbon electrodes. At these overpotentials the reduction of $Hg^{2+}$ occurs. Hence in the present investigation the reduction of $Hg^{2+}$ was expected. In the anodic path of CV any reaction was not observed.

Further, Ba and Ca are found to be electrochemically reduced at individual potentials, and the cyclic voltammogram is not reversible. Hence there is no possibility of any other reaction.

The copper was found to reduce in two steps: $Cu^+$ at early potential (-0.20 V vs. SCE) and $Cu^{2+}$ at the potential of –1.2 V vs. SCE, where the sigmoidal reduction peak was observed. The CV is reversible and the oxidation peak is observed at -0.3 V vs. SCE.

Interpretation of CV for alloy/compound electrodeposition studies is often not straightforward owing to the combined influence of a number of processes involving adsorption, electrodeposition, and coupled chemical reactions.

In the present investigation, the electrolyte (DMSO) was not involved in the electrochemical process and hence any non-significant multi-step reaction was not occurred. However, Martin-Gonzalez et al. [28] while preparing YBCO and BSCCO thick films by electrodeposition route from DMSO bath, observed the presence of characteristic bands of dimethyl sulphoxide, water and carbonates in FTIR spectra. This is due to reduction of DMSO at high deposition potential, -4.0 V vs. SCE, and was eliminated by heating the deposited samples at 100 °C. In the present investigation, all the constituents reduced well below -2.0 V vs. SCE. Hence it was ensured that DMSO is a suitable and stable solvent to proceed with.

### *4.1.2 Linear Sweep Voltammetry (LSV)*

Figure 2 shows the LSV recorded for 50 mM $HgCl_2$, and 50 mM nitrates of Ba, Ca, Cu separately dissolved in DMSO. 7 ml of each of constituents was added to get the combined bath of Hg-Ba-Ca-Cu system. The silver plate of suitable dimensions was used as substrate. The same potential range, 0 to -3.0 V vs. SCE, and scan rate of 20 mV/sec was kept during LSV measurements. From the figure 2 it can be seen that curves for bath containing Ba and Ca shows the sudden increase in current after –1.7 V and –1.9 V vs. SCE, respectively. Here, the initial flat region is due to electronic current and further sudden increase in current is due to contribution of ionic current, after reduction of constituent species. The potential at which the reduction starts is the equilibrium reduction potential or deposition potential. The relation for the current density with applied scanning potential is given by [29],

$$J = J_{iR} \text{ (electronic current)} + N Z u/V \text{ (ionic current)} \qquad (5)$$

$J_{iR}$ is the iR compensating current due to the conductivity of electrolyte (bath) and electrodes. After reduction of bath into the positive metal ions, the number of ions 'N' with valency 'Z' and velocity of ions 'u' increases with increase in applied scanning potentials and hence is called the ionic current.

The LSV for Hg showed some current at zero potential indicating that the $Hg^{2+}$ reduction starts at some positive potential range and current increases continuously in cathodic potential range. The reduction of $Cu^{2+}$ is observed at $-1.2$ V vs. SCE, and for higher potentials, curve undergoes sigmoidal nature and current decreases. It can also be noted in case of Cu that as it is a noble electro-reducing element, the magnitude of current is much higher as compared to other constituents. These large numbers of ions when reached to the working electrode, it could not be adsorbed and a double layer is developed between these metal ions. These capacitive double layers cease the ionic motion and hence current decreased.

This type of the capacitive contribution is clearly observed in LSV for Hg-Ba-Ca-Cu bath and is as shown in figure 3. Here the total observed current density during the voltammogram is expressed as the current component due to electronic current ($J_{iR}$), double layer charging ($J_{DL}$) and the charge transfer ($J_{CT}$) as [30],

$$J = J_{iR} + J_{DL} + J_{CT}$$
$$= J_{iR} + C_{DL}dE/dt + J_{CT} \tag{6}$$

where,

$C_{DL}$ is the effective double-layer capacitance.

$C_{DL}$ is generally of the order of a few microfarads per square centimeter, so its effect becomes significant for higher sweep rate $> 1$ V/sec.

This is represented in three different regions in the LSV for alloy deposition. *Region-I* represents electronic current. In *Region-II*, after electronic current, the ionic current starts and the wavy pattern results due to the dominance of individual reductions. Due to this increased ionic current, the double layer capacitance at electrode ceases the current. The individual deposition can best be obtained in this range. When the applied potential is increased further, *Region- III,* after the reduction of last species, it overcomes the capacitive barrier and current increases suddenly due to charge transfer mechanism.

Thus the knee in the curve at which the '$J_{CT}$' contribution results is considered to be the simultaneous reduction potential [31]. The linear increase in current represent the simultaneous reduction of all the constituents. Upto this potential the rate of individual deposition was ceased due to the capacitive control, and hence is said to form a capacitive controlled complex bath (of $Hg^{2+}$, $Ba^{2+}$, $Ca^{2+}$ and $Cu^{2+}$) and is reduced simultaneously after the $J_{CT}$ contribution. Hence the simultaneous reduction potential of alloy is found to be –2.0 V vs. SCE.

This is the overpotential for all the constituents in the bath and hence the condition governed by equation (3) is fulfilled. Hence potential higher than this will act as overpotential and lower potential acts as underpotential for complexing ion reduction. These potential terms greatly contribute in the kinetics of the electrodeposition process and affect the nucleation, growth and morphology of the deposit that can be controlled by selecting the appropriate potential range.

The LSV for HgBaCaCu bath, where an extra reduction peak at – 0.38 V vs. SCE was observed, was compared with the LSV of BaCaCu (figure 3) and the extra reduction peak was assigned to the possible elemental reduction of $Hg^{2+}$ at –0.38 V vs. SCE. Hence, only in $HgCl_2$ bath if at all Hg didn't reduced as $Hg^{2+}$, the 'induced deposition' of Hg might have occurred in this co-deposition process.

The induced deposition of Ca on Hg was observed by Gmelin [32]. This induced deposition would be possible due to the formation of well-known calcium amalgam where Hg and Ca catalytically co-operate with each other. Accordingly, the electrolytic separation of calcium from aqueous solution is possible only on mercury cathode by formation of amalgam. The 'Ca' reduction is clearly observed in its LSV and hence 'Hg' reduction peak in the HgBaCaCu bath could be due to 'Ca' assisted 'Hg' reduction. In the fabrication of mercury-based superconductors, formation of the non significant $CaHgO_2$ phase is observed due to amalgamation during initial thermal treatments [13].

Hence peaks in the region-II (figure 3) are attributed to the individual reduction as follows,

$$Hg^{2+} + 2e^- \rightarrow Hg \quad (-0.39 \text{ V vs. SCE})$$

$$Ba^{2+} + 2e^- \rightarrow Ba \quad (-1.30 \text{ V vs. SCE})$$

$$Ca^{2+} + 2e^- \rightarrow Ca \quad (-1.61 \text{ V vs. SCE})$$

$$Cu^{2+} + 2e^- \rightarrow Cu \quad (-0.66 \text{ V vs. SCE})$$

$$Cu^{+} + e^- \rightarrow Cu \quad (-0.20 \text{ V vs. SCE})$$

### *4.1.3 Deposition of Alloy and EDAX Studies*

It was finally decided to observe whether all species do deposit and, if so, what is the quantity of individual constituents in the final deposit? For this purpose the deposition was carried out at –2.0 V vs. SCE onto Ag substrate for 15 minutes. The film was dried at 100 °C in oxygen flowing environment for 30 minutes.

This film was then characterized by EDAX and is shown in figure 4. The characteristic peaks corresponding to Hg, Ba, Ca, and Cu have been observed. This revealed the successful reduction of all the required elements in the film and hence the co-deposition has said to be occurred. The presence of oxygen species and silver from substrate are also seen in the pattern. This is a big achievement for us towards the synthesis of Hg-Ba-Ca-CuO superconductors at low temperatures.

### *4.1.4 Formation of Complexing Bath:*

Considering the quantitative data from EDAX, individual deposition potentials and empirical electrochemical laws, an attempt was made to select the individual concentrations to get 1:2:1:2 atomic ratio. This was done with the help of LSV in view to form complexing ion bath for the controlled reduction to achieve required stoichiometry. In this approach, the bath concentrations were adjusted, nearby the required concentrations, such that the individual reduction of any of the constituents will not be dominant in the $J_{DL}$ current range. Rather, all the species must be simultaneously reduced after the potential, knee at the beginning of region-III due to $J_{CT}$. Hence many LSV's were recorded by selecting different bath concentrations nearby to the expected values, by trial and error method, and figure 5 shows LSV for one of the compositions 45 mM $HgCl_2$, and 80 mM, 65 mM and 50 mM of Ba, Ca and Cu nitrates respectively, for which the complexing reduction behaviour is observed in LSV.

Upto this, an attempt has been made to study the electrochemical behaviour of the constituents to develop the process at room temperature only. The complexing ion deposition route is belived to be worked on the basis of capacitive controlled rate of deposition.

## 4.2 Pulse Electrodeposition of HgBaCaCu Alloyed Films

The films can be co-deposited by applying either constant potential or pulsed potential. The pulsed potential deposition has many advantageous over d.c. plating technique such as; the pulse current density is considerably higher than corresponding d.c. current density which yields finer grained deposits, reduction in porosity, good control over morphology and hence improvement in mechanical and physical properties [33]. The morphology of electrodeposited materials is very important as far as superconducting transport properties are concerned. Bhattacharya *et al.* [34] have deposited TBCCO films on silver coated $SrTiO_3$ substrate both by constant potential and pulsed potential condition and found that pulse deposited films are very uniform. The critical current densities obtained are 5.6 x $10^4$ A/cm$^2$ with $T_c$ = 112 K for pulse deposited films whereas only 1 x $10^4$ A/cm$^2$ with $T_c$ = 100 K was achieved for constant potential films.

Hence in the present investigation, it was decided to deposit the HgBaCaCu alloyed films by applying pulsed potential. It is successfully demonstrated that the kinetics of the electrochemical process can be monitored by applying the pulsed potential to the complexing electrochemical bath and is presented in the following sections.

### *4.2.1 Optimization of Pulse Frequency and Duty Cycle*

Figure 6 shows the variation of cathodic current density with different frequencies of square wave pulse. For the frequencies, in the plateau region (20 Hz to 110 Hz), the cell current is due to both electronic and ionic displacements and both of these currents are flowing through the electrodeposition cell and is given by the eq. (5).

At a fixed potential 'V' and at constant temperature 'T', the velocity 'u' of the transport ion may not differ much for different ions. Hence, the time required to reach the ions from anode to cathode depends on period of applied potential V, that is, the frequency and duty cycle that considerably affect the composition and morphology of the film.

At low frequency, pulse 'ON' time is relatively greater and ions have sufficient time to reach the cathode and deposit on it. Thus, probability of formation of space charge layer increases and capacitive effect becomes important that repels further migration of positive ions and current acts as ripple DC component and hence requires greater voltage. However at high frequency, 'ON' time is very short and ions have no sufficient time to reach the cathode and deposit on it. When ions are migrated towards the cathode in the presence of electric field, they form double layer at the cathode surface. Charging time of double layer [35] is given by,

$$t_c = 17/I_p$$

and discharging time is given by,

$$t_d = 120/I_p$$

where, $I_p$ is pulsating peak current density.

When ON time (Duty cycle) is less than '$t_c$', finer grained deposits are obtained and requires larger deposition period. When 'ON' time is greater than '$t_c$', it acts like ripple DC. Hence, the optimum condition is given as [36],

*Pulse 'ON' time < $t_c$ and Pulse 'OFF' time > $t_d$*

Pawar *et al.* [37] have studied electrical properties of Bi-Sr-Ca-CuO superconducting thin films deposited at 25 Hz and at different duty cycle, and found that the $T_c$ varies with duty cycle and was maximum for 50 % duty cycle. Hence, DC pulsating potential with 25 Hz frequency and 50 % duty cycle was applied between the electrodes.

### 4.2.2 Deposition Potential and Current Density

Here the same three-electrode cell and the finally optimized bath was used and potential was applied by pulse generator. The polarization curve was plotted to determine the complex ion deposition potential for the deposition of HgBaCaCu alloy onto Ag substrate and is shown in figure 7. It is seen that the complex ion reduction current increases at about – 1.7 V vs. SCE. This potential is less than that obtained by DC potential because of the fact that due to 'OFF' time pulsating potential allows to discharge the double layer relatively in a small time period and also the peak current density at peak potential is higher than corresponding DC current.

The actual deposition potential was estimated by studying the deposition current density (chronoamperometry) for different cell potentials. Figure 8 shows the deposition current density recorded at the cell potentials of -1.8 V, -1.7 V and -1.6 V vs. SCE. From the figure it is observed that deposition current density sharply decreases in first few seconds and then remains steady. In pulse electrodeposition, during pulse 'ON' time the corresponding peak current is higher than DC potential. Rapid fall of current is observed in first few seconds and it is attributed to the formation of double layer between electrode-electrolyte interfaces, which causes an increase in resistance. Whereas during pulse 'OFF' time these migrated metal ions get sufficient time to deposit, and hence the nucleation process increases. Thus, after few seconds the current density remains almost steady. In case of the deposition potential of –1.8 V vs. SCE; however, the current density gradually increases rather than the steady state value. This increase in current density results in the formation of non-uniform and spongy deposit. In this type of current density, the further increase in current may be due to the catalytic activities of any of the constituents or if concentration of any of the constituents decreases then the current density of next electroactive species increases faster and hence, over the period, the deposit ratio could not be ensured. If this is not the case and if the stoichiometry remains, according to the kinetics of the nucleation, the nature of nucleation may be instantaneous and three-dimensional deposition occurs [30].

The current density at –1.7 V vs. SCE is constant over the time representing the steady flow of ions and this control may be due to the controlled pulsating current and minimized interface double layer capacitance. According to the kinetics of the electrodeposition governed by the Buttler-Volmer equations for the nucleation [38], this type of current density nature results into progressive three-dimensional growth. Here current density is relatively low and steady than at –1.8 V vs. SCE, this gives uniform, fine-grained non-porous deposits. The same type of current density is observed at –1.6 V vs. SCE but due to further decreased current density the granules size increases and also required longer time for higher thickness.

Hence potential of –1.7 V vs. SCE was selected as the optimum deposition potential for Hg-Ba-Ca-Cu alloy.

*4.2.3  Thickness of Alloyed Film*

Figure 9 shows the variation of thickness with deposition time for Hg-Ba-Ca-Cu film deposited at –1.7 V vs. SCE. Films were dried at 100 $^o$C for half hour to remove moisture and DMSO in the deposit. It has been observed that the thickness increases linearly with time, as there is steady flow of ions from anode to cathode. As the thickness increases with time, the deposition continues either by build-up on previously deposited material i.e. on old nucleation centers or the formation and growth of new ones takes place. However, after 30 minutes of deposition the films were found to be spongy and for further increase in deposition period deposits peels off in the electrolyte. The optimum thickness of about 2 μm was obtained for 20 min deposition.

### *4.2.4 Current Time Transient and Nucleation Growth*

The chronoamperometry, the variation of current with time, was measured for the complexing alloy bath at a potential of –2.0 V vs. SCE (see section 4.1.4) using the VersaStat-II and is as shown in figure 10 (a).

To ascertain whether the nucleation process is instantaneous or progressive and what is the structure of final growth, it is more convenient to represent the potentiostatic transients in a dimensionless plot of $(I/I_{max})^2$ vs. $(t/t_{max})$, where $I_{max}$ and $t_{max}$ corresponds to the maximum current density and the corresponding time in the transient. The relation between $(t/t_{max})$ and $(I/I_{max})^2$ is given by [39],

$$(I/I_{max})^2 = 1.9542 (t_{max}/t) \ [\ 1-\exp\{-1.2564(t/t_{max})\}]^2 \ \text{--------Instantaneous}$$

$$(I/I_{max})^2 = 1.2254(t_{max}/t) \ [\ 1-\exp\{-2.3367(t/t_{max})^2\}]^2 \ \text{--------- Progressive}$$

The observed data is fitted with the theoretical plots for instantaneous and progressive growth and is shown in figure 10 (b).

The figure 10 (b) shows that the obtained transient fits to the progressive growth. In a current time transient, current before approaching the steady state value passes through a single maximum and three dimension growth results.

Depending on the time span of this maxima and the steady state, recently, the theoretical model is developed by M.Y. Abyanesh [40] and predicted the triangular and hemispheroidal growth habits and are shown in figure 10 (c) for the representation. Our experimental results (Figure 10 (b)) also show such current time transient suggesting the three dimensional progressive growth occurs after steady state value leading to the growth of hemispherical nuclei.

Figure 10 (d) shows the SEM (10000 x) where the hemispheroidal granules can be seen. This current time transient model could help in predicting the possible growth habits without looking at the microstructural measurements and hence growth could be *in-situ* monitored by applying the suitable potential range and texture of the substrate.

Here, as the data is to collect within first few seconds the constant potential was applied using VersaStat-II and electrochemistry software was used to take the real time measurements.

## 4.3 Electrochemical Oxidation of HgBaCaCu Alloy

It is for the first time that the HgBaCaCu alloyed precursor films are fabricated in co-deposition route at room temperature. In order to convert this material into superconducting phase, $HgBa_2Ca_1Cu_2O_{6+\delta}$, it is required to add oxygen crucially to maintain the oxygen stoichiometry in the sample. The problem of volatile and toxic nature of mercury and mercury oxide at high temperature oxidation is widely addressed earlier. Thus it was decided to employ the novel electrochemical oxidation route that works at room temperature.

The conventional three-electrode system was used with HgBaCaCu precursor deposited onto Ag as a working electrode. The alkaline 1 N KOH solution in double distilled water was used as electrolyte bath to get oxygen species in anodic potentials. The figure 11 shows the LSV recorded both onto Ag and Ag/HgCaCaCu electrodes at the potential range of 0 to +1.0 V vs. SCE. For Ag, the peak at + 250 mV is seen which is assigned to the reaction,

$$KOH\ (aq) \rightarrow K^+ + OH^-$$

Further broad peak at + 630 mV vs. SCE is assigned to the reaction at anode,

$$4OH^- \rightarrow O_2\uparrow + 2H_2O + 4e^-$$

After some over potential, 700 mV, the current increases fast resulting in oxygen evaluation at silver electrode.

The nature of LSV on Ag/HgBaCaCu alloy electrode is similar with some shifts in the peaks in anodic range. This shift may be due to low conductivity of working electrode after deposition of HgBaCaCu alloy.

From these observations it was decided to oxidize the alloyed HgBaCaCu films at +0.7 V vs. SCE.

A typical chronoamperometric plot for the electrochemical oxidation of HgBaCaCu film at +0.7 V vs. SCE is shown in figure 12. Initially current decreases steeply and then slightly increases for some time representing diffusive or intercalating nature and finally decreases slowly with respect to time. The available oxygen sites are filled up within first few seconds only and hence the current decreases, as the further intercalation is not possible. However, considering Nernst equation, one can estimate this oxygen activity to be equivalent, at room temperature for a few hundred mV of over potential, to a very high oxygen pressure. Hence by oxidizing the samples for further lengths of time it could be possible to intercalate extra non-stoichiometric oxygen content and even further over oxidation of the samples as can be seen from the following reaction [41].

$$La_2CuO_4 + 4\delta OH^- \longrightarrow La_2CuO_{4+\delta} + 2\delta H_2O + 4\delta e^-$$

In our case,

$$Hg_1Ba_2Ca_1Cu_2O_6 + 6\delta OH^- \longrightarrow Hg_1Ba_2Ca_1Cu_2O_{6+\delta} + 3\delta H_2O + 6\delta e^-$$

Hence in the present investigation the HgBaCaCu alloyed samples were oxidized for different lengths of time and its structural and electrical properties were measured.

## 4.4 Variation of Structural and Electrical Properties with Oxidation Period :

### *4.4.1 X-Ray Diffraction (XRD) Studies:*

The alloyed films were oxidized electrochemically for different durations and their XRD patterns were recorded. Figures 13 (a to e) show the XRD patterns for a) As-deposited, b) 20 minutes, c) 24 minutes, d) 28 minutes and e) 30 minutes electrochemically oxidized films, respectively. The XRD pattern of as-deposited film shows non-stoichiometric crystalline growth representing the pulsed-electrocrystallization of HgBaCaCu alloy. Some oxide phases are found to be developed which may be due to the exposure of the precursor to the atmosphere. These are noted to be of $CaHgO_2$ at $2\theta = 18.2^o$ and $32^o$ [42-47]. The film oxidized for 20 minutes shows the development of crystallization and indexed with tetragonal indices with *P4/mmm* space group for Hg-1212 phase [48]. The film was polycrystalline in nature with presence of (002), (003), (102), (004), (111), (005), (104), (114), (115), (200) and (107) planes. It contains some impurity phase of $CaHgO_2$. The lattice parameters were calculated and are $a = 3.872$Å and $c = 12.648$ Å. These values do not matched adequately with the standard values [49]. Hence another sample was further oxidized for 24 minutes and it was observed that impurity phases are diminished and crystallinity is improved by further oxygenation. This is estimated by measuring the lattice parameters. The films were further oxidized for 28 minutes and XRD data showed the presence of single phase Hg-1212 with improved crystallinity and lattice parameters are in agreement with standard values. The film oxidized for 30-minute show presence of some oxide impurities. The Table 1 shows the standard and calculated lattice parameters of these Hg-1212 films for different durations of oxidation.

From the Table 1 it is seen that the lattice parameter *a* decreases with increase in oxidation period. The variation of *a* and *c* with oxidation period is shown graphically in Figure 14.

Variation in lattice parameters clearly indicates the intercalation of oxygen species. The presence of impurity phases might be due to insufficient oxygen content, which then converts into superconducting state after further oxygenation. The optimum oxidation period of 28 min. was also confirmed by superconducting transition temperature reported in section 4.4.3.

*4.4.2 Microstructural Studies:*

Figure 15(a-d) shows the SEM of as-deposited, 20 minute, 24 minute, 28 minute and 30 minute oxidized films. It is seen that the films are uniform and dense. The as-deposited films have circular hemispherical growth. In pulse plating, the pulsed current density remains considerably higher with time, which leads to a higher population of adatoms on the surface during pulse deposition resulting in an increased nucleation rate and therefore a uniform and finer grained structure.

But the film oxidized for 20 minutes showed a noticeable change in surface morphology. The granular structure with fine grain is observed. The grain growth and compactness was improved for the films oxidized for 24 minutes and 28 minutes. When oxygen is intercalated, due to the intercalating channels the granularity is developed and its size was found to be increased for further oxidation periods. The improved grain size is attributed to the intercalation of oxygen and homogenization of the identical phases in a granular depending on the initial growth habits of the critical nucleii. The polycrystallinity may be due to such a granular growth.

*4.4.3 Temperature Dependence of Electrical Resistivity:*

Figure 16 shows the variation of normalized electrical resistance of HgBaCaCuO films electrochemically oxidized for different periods. The as-deposited film showed metallic behaviour and do not became superconducting still up to 77 K. Upto 20 min. of electrochemical oxidation the films were not found to be superconducting throughout the temperature range of measurement. This is attributed to the underdoped state of the samples. But the film oxidized for 20 minute turned superconducting at 87 K. Further, the films oxidized for 24 minutes and 28 minutes showed increase in superconducting transition temperature to be 97 K and 104.7 K, respectively. The film oxidized for 30 minutes showed transition temperature $T_c$ = 101 K, which is less than the films oxidized for 28 minutes. This decreased $T_c$ might be due to the over oxygen content of the sample. As the oxidation period is increased the amount of extra non-stoichiometric oxygen content in $HgO_\delta$ layer increases which intern increase the hole concentration in $CuO_2$ planes. The transition temperature is found to be lower and transition widths are found to be broad. The onset superconducting transition temperature shows the formation of $HgBa_2CaCu_2O_{6+\delta}$ films.

### *4.4.4 Critical Current Densities ($J_c$):*

The critical current densities, $J_c$, were recorded with the criteria of maximum current ($I_c$) that can be passed to develop the potential gradient of 1μV/cm. The values recorded are listed in Table 2. It is seen that as oxygen content increases $J_c$ values increases representing the complete stoichiometric oxidation of the samples. However, the values are lower compared to the reports [43]. This may be due to the weak link behaviour at the grain boundary.

## 5. Discussion

Here, in the electrochemical deposition of the films, according to the established atomistic model [50], the applied potential allows the metal ions in liquid phase to transform its phase to solid and adsorbs onto the substrate simply by following the reaction $M^{2+} + 2e^- \rightarrow$

M. Single atom or group of atoms constituting alloy/compound adsorbed on the active site can be considered as a stable cluster which grows spontaneously and formally the active sites itself plays the role of critical nucleii. In this case the thermodynamic work for the nucleus formation equals zero and therefore only the kinetics determine the rate of phase formation [51]. Here the overall reaction takes place on the atomic level and it reduces the time and temperature that is otherwise required for the formation of cuprate superconductor by powder calcining approach. Hence this principle was attempted to workout the process for HgBaCaCuO formation at room temperature.

The kinetics of the nucleation and control over the morphology is achieved by employing pulsed potential and allowing the double layer to discharge in a controlled way, which results in the formation of smooth uniform films. The optimized process parameters for the deposition of HgBaCaCu alloy films are,

- Electrochemical Bath :   45 mM $HgCl_2$, 80 mM $Ba(NO_3)_2$, 65 mM $Ca(NO_3)_2.3H_2O$, 50 mM $Cu(NO_3)_2.2H_2O$ in Dimethyl Suplhoxide (DMSO)
- Pulse Parameters:   Frequency :- 25 Hz, Duty Cycle :- 50 %
- Deposition potential:   -1.7 V   vs. SCE
- Thickness:   2-3 micrometers
  For 20 minutes it is 2 $\mu m$

The pulse frequency and duty cycle of pulsed potential was so optimized that after 5 minutes of deposition the films looks to be well covered with the substrate. The current density was steady representing that ion flow from anode to cathode and hence the charge transfer at electrode-electrolyte interface is uniform and hence the thickness increases gradually with respect to time. Hence it is possible to obtain the desired HgBaCaCu alloyed film with desired thickness by varying the length of deposition period.

The silver substrate is found to be suitable to provide the sufficient active sites for the nucleation growth. But as the substrate itself is polycrystalline and there is a lattice mismatch of 5.95% between Ag and Hg-1212 unit cell, the polycrystalline growth is observed. However, by using the textured substrates i.e by depositing the silver thin layer onto oriented LiAlO$_2$ or SrTiO$_3$, Bhattacharya *et al.* [52] achieved the *c*-axis oriented films of TBCCO with $J_c$ values greater than 10$^5$ A/cm$^2$. Martin-Gonzalez *et al.* [53] have deposited the YBCO films onto the non-textured silver foils and observed that the 30 micron thick film cracks when heated to the 700 $^o$C. However, in the present investigation we have not deposited the thick films and did not heat-treated also and hence no cracks in the samples are seen.

The use of DMSO is found to be significant, otherwise in most of the studies, where aqueous electrolyte used was found to develop the impurity phases. The reaction takes place as,

$$M^{z+} + xH_2O + ze^- \rightarrow [M] + x\,H_2O,$$

and when the electrolyte gets polarized the hydroxyl ions form,

$$M + 2OH \rightarrow M(OH)_2$$

and hence there is a formation of Ca(OH)$_2$, Ba(OH)$_2$, Cu(OH)$_2$ [53].

These can be heat treated ($\sim$ 600 $^o$C) and converted to the respective oxides to form the constituting superconducting films but it is difficult to control oxygen stoichiometry besides the formation of impurity phases such as Cu$_2$O, BaCuO$_2$ etc. Mercury evaporation takes place at such high temperatures. Otherwise it is a best method for the oxide formation.

Hence, in the present investigation, the controlled electrochemical oxidation of alloyed film was carried by using alkaline KOH bath. The electrochemical intercalation of oxygen into $La_2CuO_4$ using anhydrous organic electrolyte media of DMSO/0.1 M $NaClO_4$ or $nBu_4NBF_4$ with $KO_2$ as oxygen source was used by Jacob *et. al.* [54], and the sample was oxidized which show superconductivity at 42 K. But the duration needed (76 hrs) is large and this long polarization time induces degradation and corrosion of materials due to dissolution-precipitation process during the reaction.

But in our experiment the desired optimum oxygen content was achieved within 28 minutes at room temperature. This may vary with the bulk thickness.

Figure 17(a) shows the transition temperature as a function of electrochemical oxidation period. This cupola like behaviour is similar as observed by Fukuoka [55], Antipov [56] and Paranthaman [57] for the $T_c$ of Hg-1212 as a function of $\delta$. Accordingly, oxidation period of 28 minute for alloyed film with 2μm thickness is revealed to be optimum.

The representative figure 17(b) of Fukuoka's work is shown here for the comparison. It is seen that optimum $\delta$ calculated by different groups is different but the superconducting nature of under-doped, optimum and overdoped states are same. The $\delta = 0.22$ [49] is considered to be the optimum because the oxygen content in this case is measured by iodometric titration and powder neutron diffraction method and found to be same, with $T_c = 127$ K.

Figure 14 shows variation of lattice parameters *a* and *c* with electrochemical oxidation period. It can be seen that *a* parameter decreases with oxygenation and shows the cupola like behaviour in $T_c$ (inset of Figure 14). This decrease in $T_c$ results due to the fact that when O3 (HgO) site is occupied the Hg-$O_\delta$ bond length reduces. At the same time this induced oxygen induces the hole concentration in $CuO_2$ layers and copper valence increases (+2.05 to +2.25) and the $CuO_2$ layer slightly contracts and Cu1-O1-Cu1 becomes flat [59].

Similarly, it is seen that as oxidation period increases the *c* increases with increase in $T_c$. This also reveals the increase in $\delta$ with oxidation period. As predicted before [57], as $\delta$ is increased to optimum value the 'Ba' atoms in BaO plane interacts with oxygen at O3 site and looses the interaction with oxygen in $CuO_6$ (Cu1-O1) octahedra. This results in increasing the bond length of Cu1-O2. Overall, the $HgO_2$ units reside farther from the $CuO_2$ plane. Hence *c* parameter might have increased. It was found to be in close agreement for 28-minute oxidation period, which is revealed to be optimum. These features of lower in-plane Cu1-O1 bond length and its flatness (*a*-parameter); and longer Cu1-O2 (*c*-parameter) in mercury superconductors are considered to be the reason for bearing the higher transition temperatures than thallium based superconductors [59].

Hence on the basis of structural parameters it is concluded that the Hg-1212 phase is successfully formed. But the superconducting transition temperature $T_c$ and $J_c$ values are lower. Particularly, the transition width is found to be maximum as given in the Table 4.

This broad transition might be due to the weak links caused by the grain boundary, anisotropy contributions due to polycrystalline growth. This $T_c$ and $J_c$ values could be improved by controlled heat treatments below the decomposition of mercury and mercury oxides.

## 6. Conclusions

It is concluded that the complexing ion simultaneous deposition of HgBaCaCu films have been successfully carried out by pulse electrodeposition techniques at room temperature. The deposition potential of –1.7 V vs. SCE was the over potential for the reduction of all individual constituents, those deposits simultaneously and atomic ratio (Hg-1212) is monitored by controlling the concentrations of constituents in the bath.

The room temperature electrochemical oxidation route is proved to be a novel technique to intercalate oxygen species into the lattice of these alloyed films where oxygen content can be monitored by varying the period of oxygenation and is revealed by the variation of $T_c$ ranging from 87 – 104.7 K and $J_c$ from 830 - 1780 A/cm$^2$.

In pulse electrosynthesis technique, the reaction takes place on atomic scale and it reduces temperature and time of the process. As it avoids the high temperature processing, it prevents from inhalation of toxic mercury oxide vapors those are harmful to human being. Hence, this is environmental friendly and economical method for synthesis of superconducting thin films with required stoichiometry. None of the other processes could fabricate the Hg based superconducting films at room temperatures.

**Acknowledgements**

Authors wish to thank the University Grants Commission, New Delhi (India), for financial support under superconductivity R & D Project and Dr. A.V. Narlikar for his constant encouragement. One of the authors, DDS, thanks Council of Scientific and Industrial Research (CSIR), New Delhi for the award of Senior Research Fellowship.

**Figure Captions**

Figure 1   Cyclic Voltammetry curves for (a) DMSO solvent and 50 mM solutions of (b) $HgCl_2$, (c) $Ba(NO_3)_2$, (d) $Ca(NO_3)_2.2H_2O$ and (e) $Cu(NO_3)_2.3H_2O$ into DMSO

Figure 2   Linear Sweep Voltammetry (LSV) for 50 mM solutions of
(a) $HgCl_2$, (b) $Ba(NO_3)_2$, (c) $Ca(NO_3)_2.2H_2O$,
(d) $Cu(NO_3)_2.3H_2O$ and (e) Combined HgBaCaCu bath into DMSO on Ag substrate

Figure 3   LSV for the bath of 50 mM concentrations of Hg, Ba, Ca and Cu constituents together.

Figure 4   EDAX pattern for the HgBaCaCu film deposited from 50 mM bath concentrations of each constituents.

Figure 5   LSV for the bath containing 45 mM $HgCl_2$ and 80 mM, 65 mM and 50 mM of Ba, Ca and Cu nitrates, respectively, forming a complexing bath in DMSO.

Figure 6   Variation of cathodic current density with pulse frequency

Figure 7   Polarization curve for the HgBaCaCu alloy deposition

Figure 8   Variation of cathodic current density with deposition time during deposition of Hg-Ba-Ca-Cu alloyed film for different potentials with SCE electrode

Figure 9   Variation of thickness with deposition time of the Hg-Ba-Ca-Cu alloyed films

Figure 10   (a) Chronoamperometry curve for the deposition of HgBaCaCu onto Ag substrate
(b) The fitting of observed transient with theoretical curves of instantaneous and progressive growths.
(c) The theoretically predicted 1) triangular and 2) hemispherical nucleation growth.
(d) SEM of the HgBaCaCuO film.

Figure 11   LSV for the electrochemical oxidation using alkaline 1N KOH solution onto
(a) Ag substrate and
(b) HgBaCaCu alloyed film deposited on Ag substrate

Figure 12   Variation of current density with time (chronoamperometry) during electrochemical oxidation of HgBaCaCuO alloyed film

Figure 13   The X-ray diffraction pattern for (a) as-deposited HgBaCaCu alloyed film and (c) 20 min.; (c) 24 min.; (d) 28 min.; and (e) 30 min. electrochemically oxidized films

Figure 14   Variation of *a* and *c* parameter with oxidation period.

Figure 15   Scanning electron micrographs for
(a) as-deposited HgBaCaCu alloyed film and electrochemically oxidized films for (b) 20 min.; (c) 24 min.; (d) 28 min.; and (e) 30 min.

Figure 16   Temperature dependence of normalized resistance of the Hg-Ba-Ca-CuO (Hg-1212) films electrochemically oxidized for different periods

Figure 17   (a) Transition temperatures achieved for the films oxidized for different electrochemical oxidation periods.
(b) Variation in $T_c$ with δ as observed by Fukuoka [55].

**Tables**

*Table 1. The standard and calculated lattice parameters of the Hg-1212 films.*

| Oxidation period → Lattice parameters ↓ | Standard [49] | 20 min. | 24 min | 28 min | 30 min |
|---|---|---|---|---|---|
| $a$ (Å) | 3.8580 | 3.872 | 3.867 | 3.8576 | 3.8559 |
| $c$ (Å) | 12.6811 | 12.648 | 12.650 | 12.658 | 12.663 |

*Table 2. $J_c$ values of the Hg-1212 films electrochemically oxidized for different durations.*

| Oxidation period (min.) | $J_c$ A/cm$^2$ |
|---|---|
| 20 | 830 |
| 24 | 1120 |
| 28 | 1437 |
| 30 | 1780 |

*Table 3. The structural parameter, extra oxygen content δ and $T_c$ for some HgBa$_2$CaCu$_2$O$_{6+\delta}$ samples*

| REF. | $a$ (Å) | $c$ (Å) | $T_c$ (K) | δ | Cu-O1 (Å) | Cu-O2 (Å) |
|---|---|---|---|---|---|---|
| 58 | 3.85766 | 12.6562 | 120 | 0.265 | 1.936 | 2.753 |
| 57 | 3.8570 | 12.6923 | 122 | 0.33 | 1.9285 | 2.81 |
| 56 | 3.8543 | 12.6416 | 123 | 0.28 | 1.9272 | 2.7986 |
| 59 | 3.8601 | 12.7030 | 126 | 0.21 | 1.9301 | 2.825 |
| 60 | 3.8580 | 12.681 | 127 | 0.22 | 1.9290 | 2.801 |

*Table 4. Variation in transition width with electrochemical oxidation period.*

| Oxidation period (min.) | $T_c^{onset}$ (K) | $T_c^0$ (K) | Transition width (K) |
|---|---|---|---|
| 20 | 107 | 87 | 20 |
| 24 | 110 | 98 | 12 |
| 28 | 114 | 104.7 | 9.3 |
| 30 | 114 | 101 | 13 |

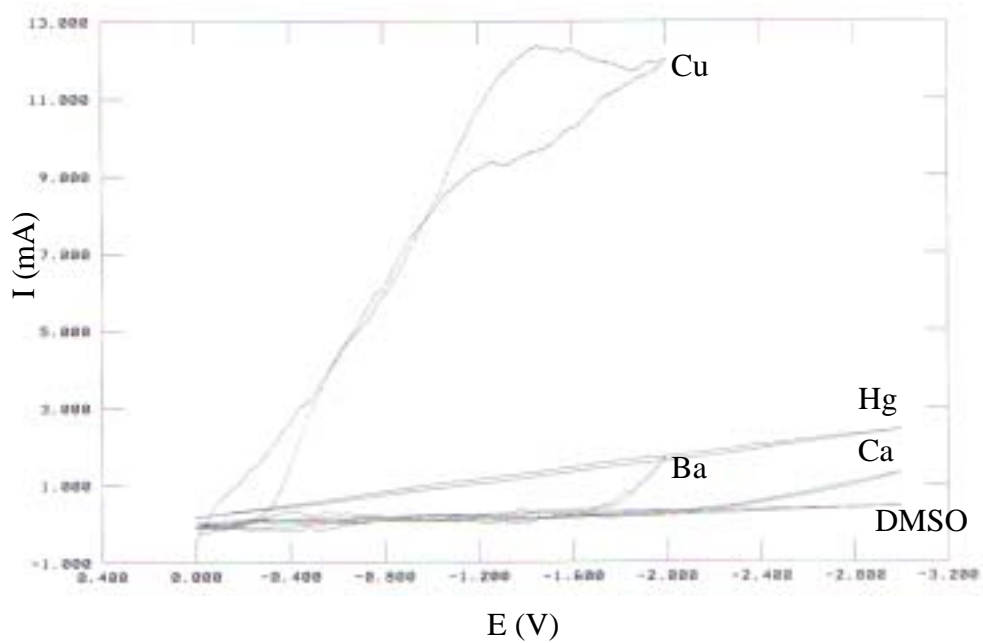

Figure 1  Cyclic Voltammetry curves for (a) DMSO solvent and 50 mM solutions of (b) $HgCl_2$, (c) $Ba(NO_3)_2$, (d) $Ca(NO_3)_2.2H_2O$ and (e) $Cu(NO_3)_2.3H_2O$ into DMSO

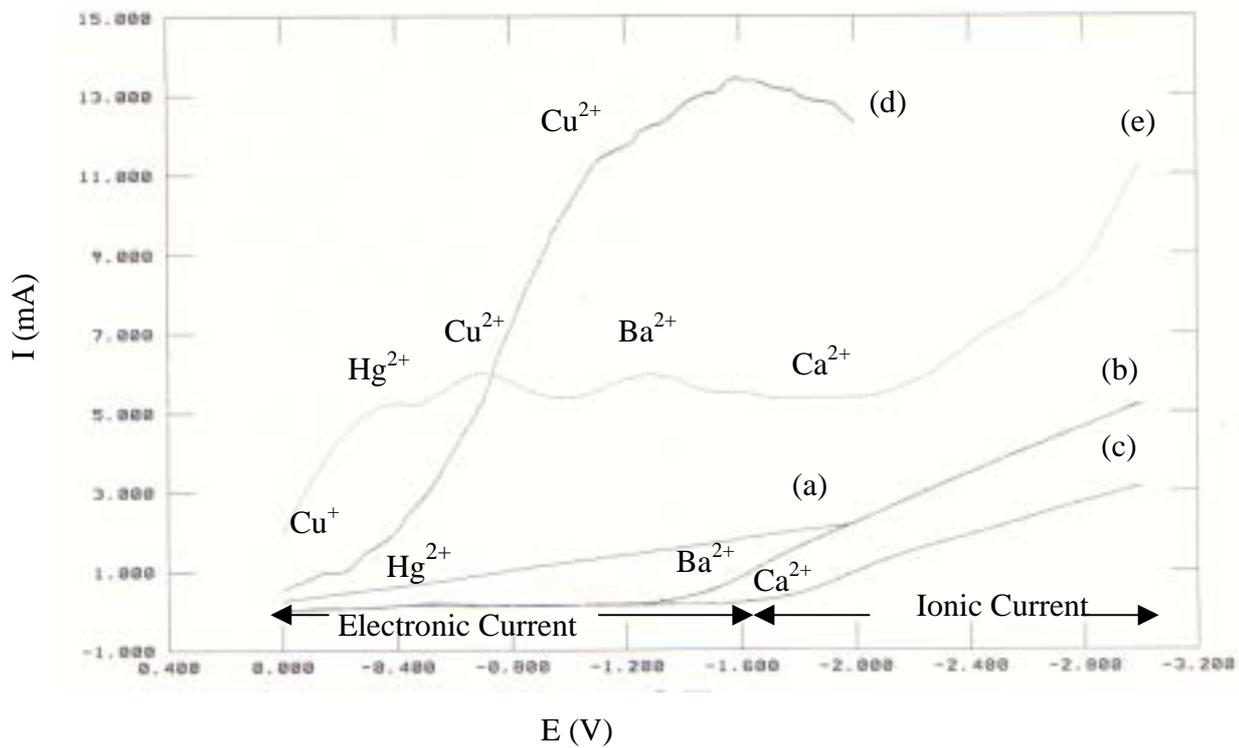

Figure 2  Linear Sweep Voltammetry (LSV) for 50 mM solutions of
(e) HgCl$_2$, (b) Ba(NO$_3$)$_2$, (c) Ca(NO$_3$)$_2$.2H$_2$O,
(d) Cu(NO$_3$)$_2$.3H$_2$O and (e) Combined HgBaCaCu bath into DMSO on Ag substrate

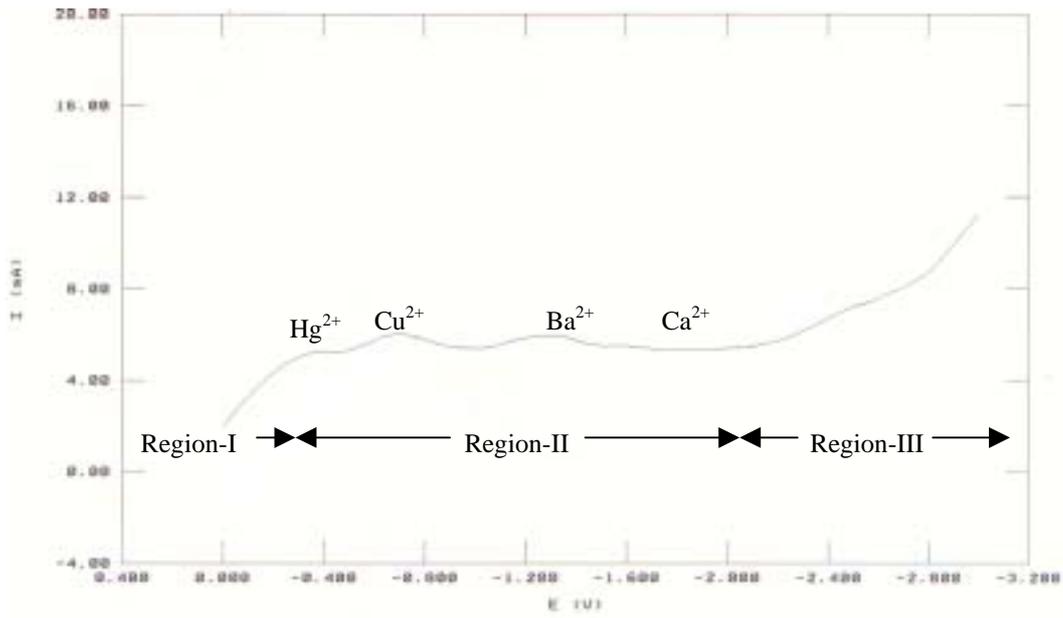

Figure 3. LSV for the bath of 50 mM concentrations of Hg, Ba, Ca and Cu constituents together.

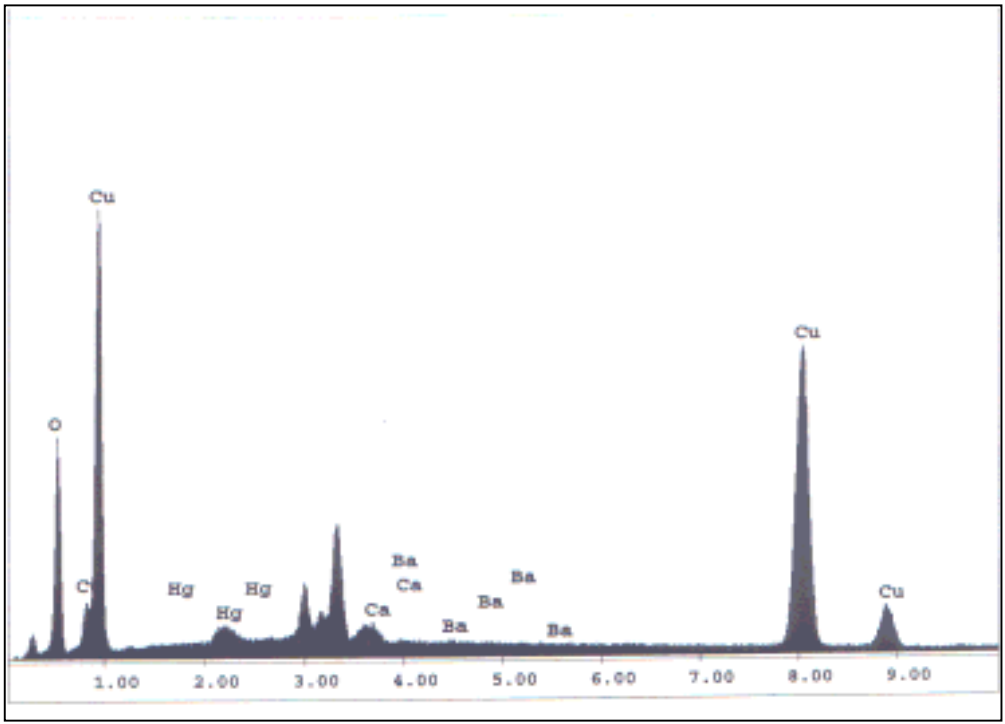

Figure 4 : EDAX pattern for the HgBaCaCu film deposited from 50 mM bath concentrations of each constituents.

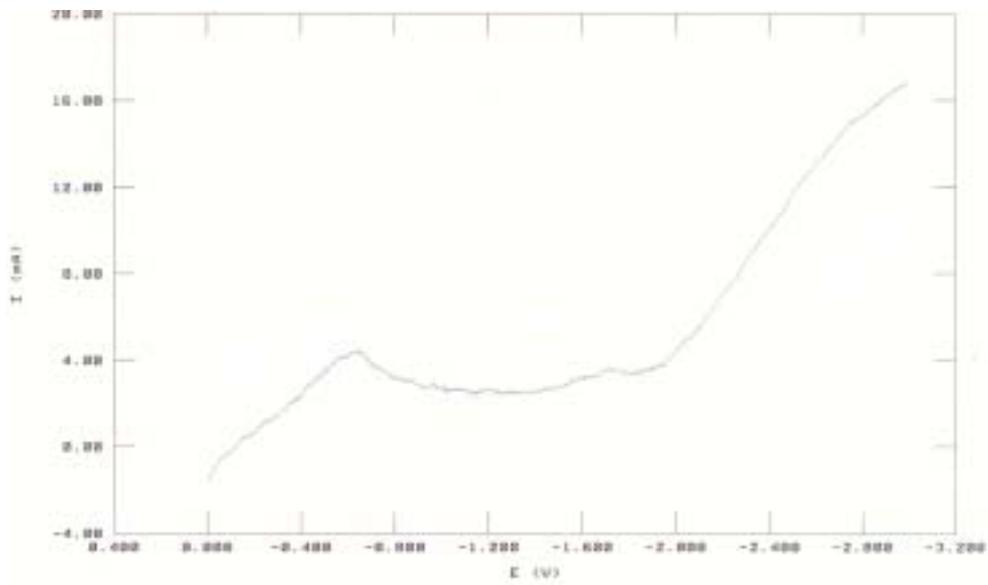

Figure 5   LSV for the bath containing 45 mM $HgCl_2$ and 80 mM, 65 mM and 50 mM of Ba, Ca and Cu nitrates, respectively, forming a complexing bath in DMSO.

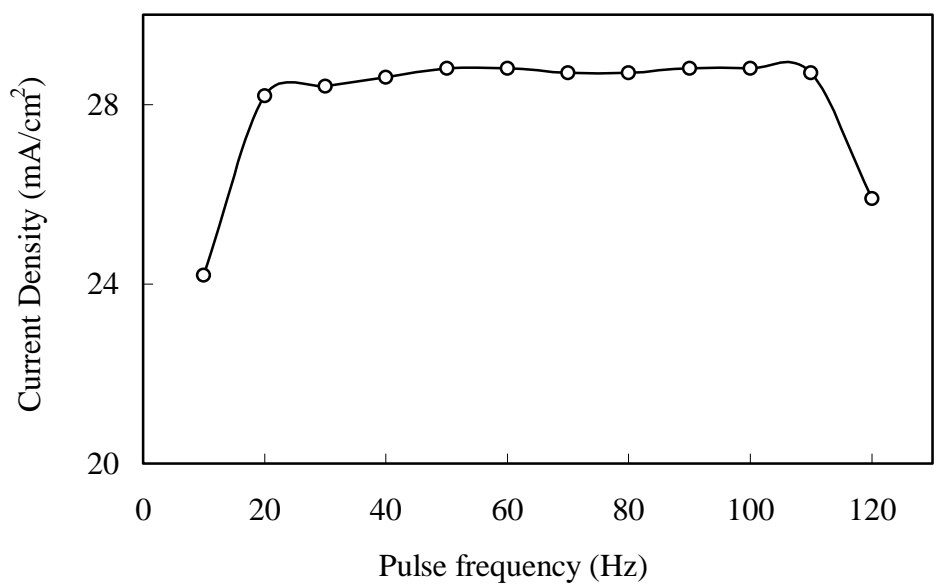

Figure 6   Variation of cathodic current density with pulse frequency

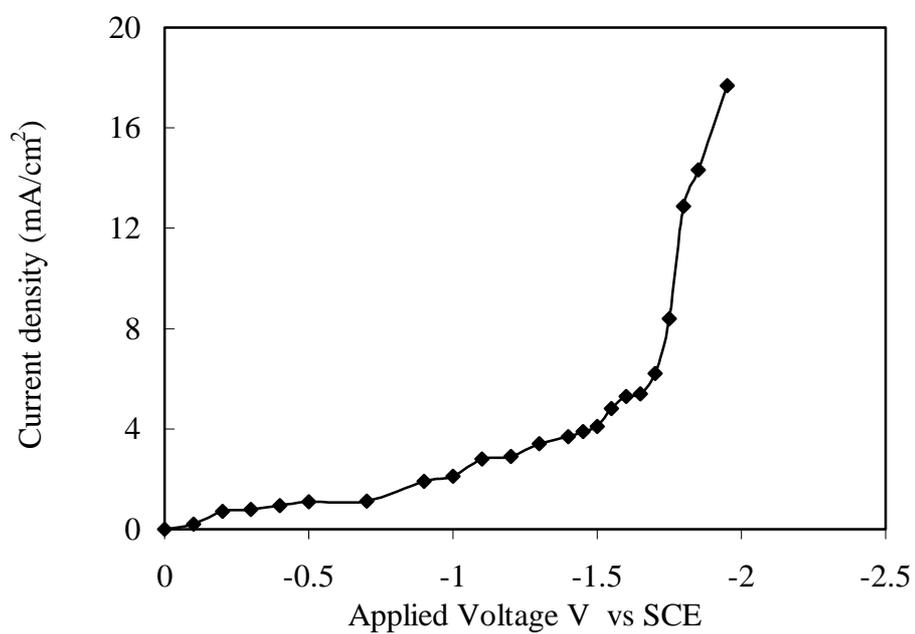

Figure 7   Polarization curve for the HgBaCaCu alloy deposition

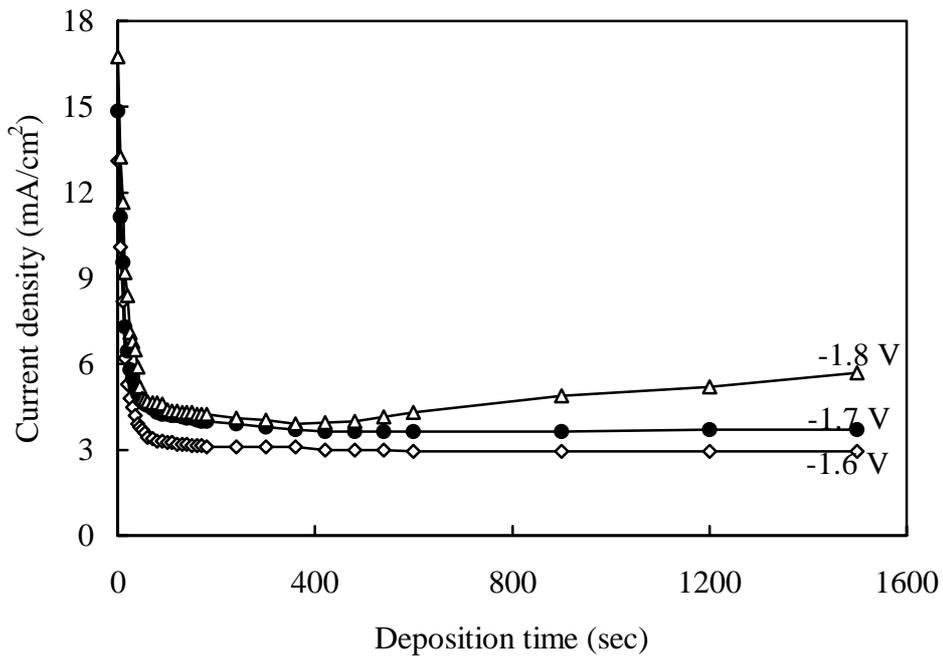

Figure 8 Variation of cathodic current density with deposition time during depostion of Hg-Ba-Ca-Cu alloyed film for different potentials with SCE electrode.

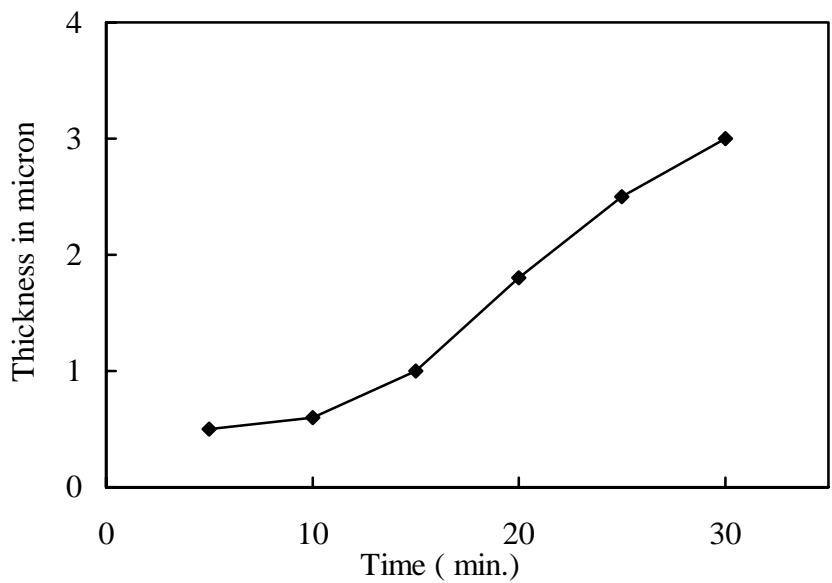

Figure 9. Variation of thickness with deposition time of the Hg-Ba-Ca-Cu alloyed films

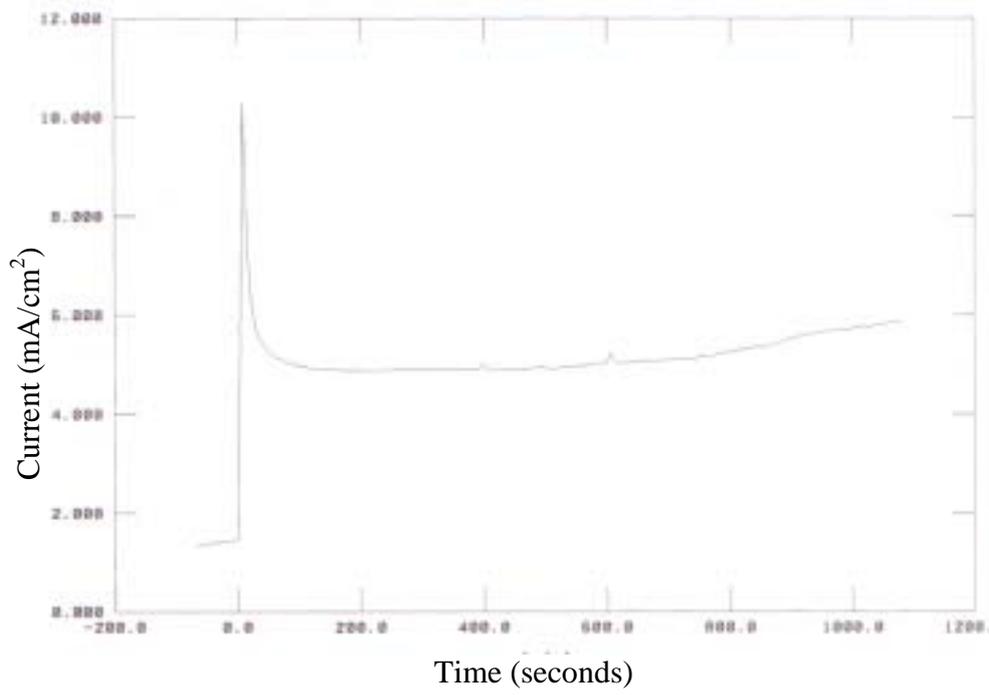

Figure 10 (a) Chronoamperometry curve for the deposition of HgBaCaCu onto Ag substrate

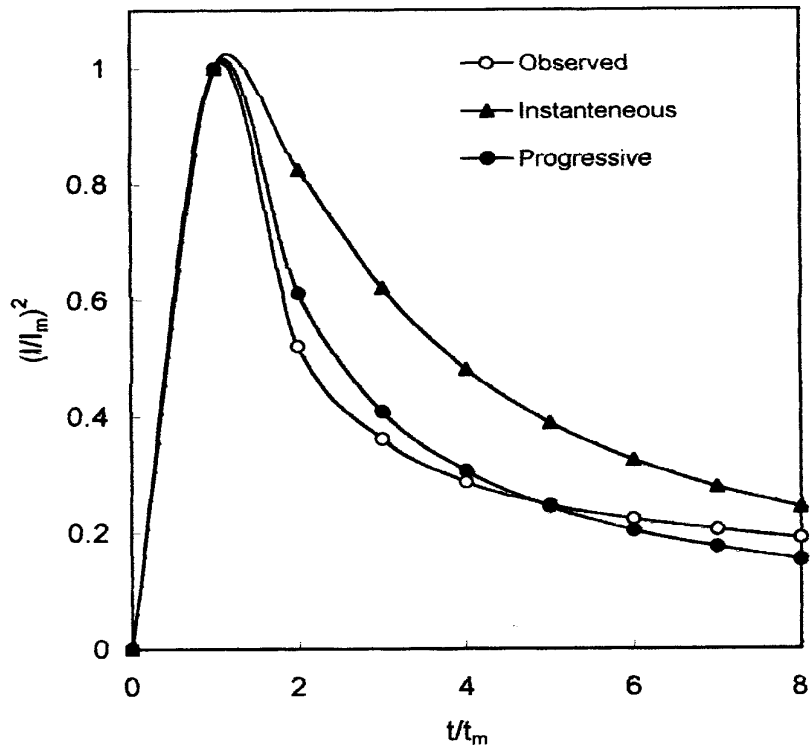

Figure 10 (b). The fitting of observed transient with theoretical curves of instantaneous and progressive growths.

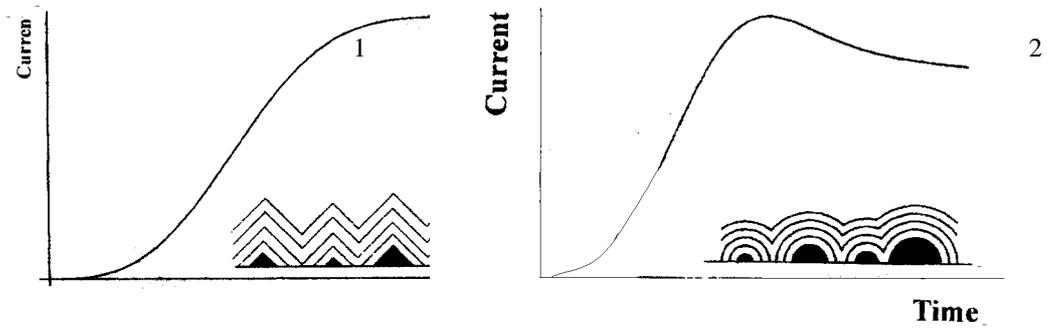

Figure 10 (c). The theoretically predicted 1) triangular and 2) hemispherical nucleation growth.

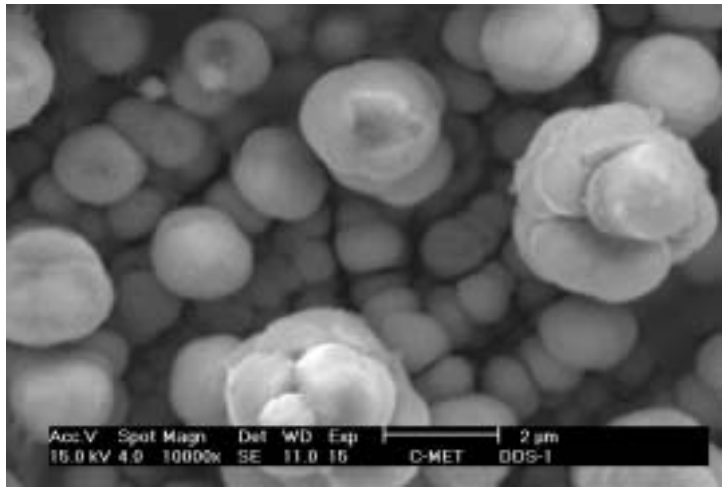

Figure 10(d) SEM of the HgBaCaCuO film

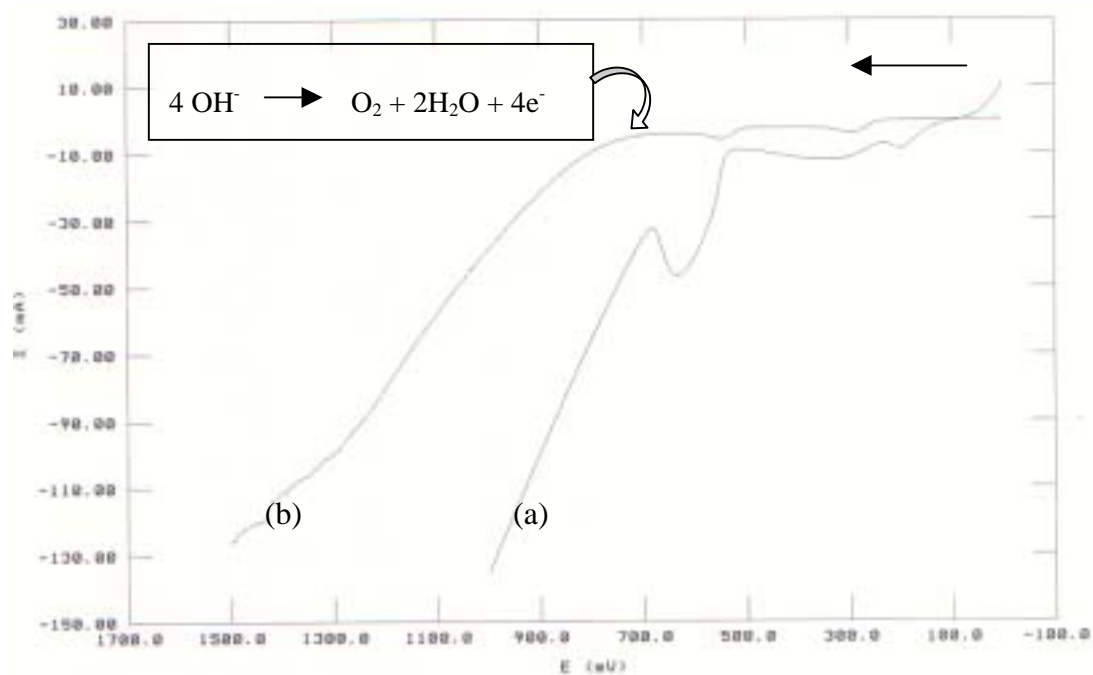

Figure 11: LSV for the electrochemical oxidation using alkaline 1N KOH solution onto (a) Ag substrate and
(b) HgBaCaCu alloyed film deposited on Ag substrate

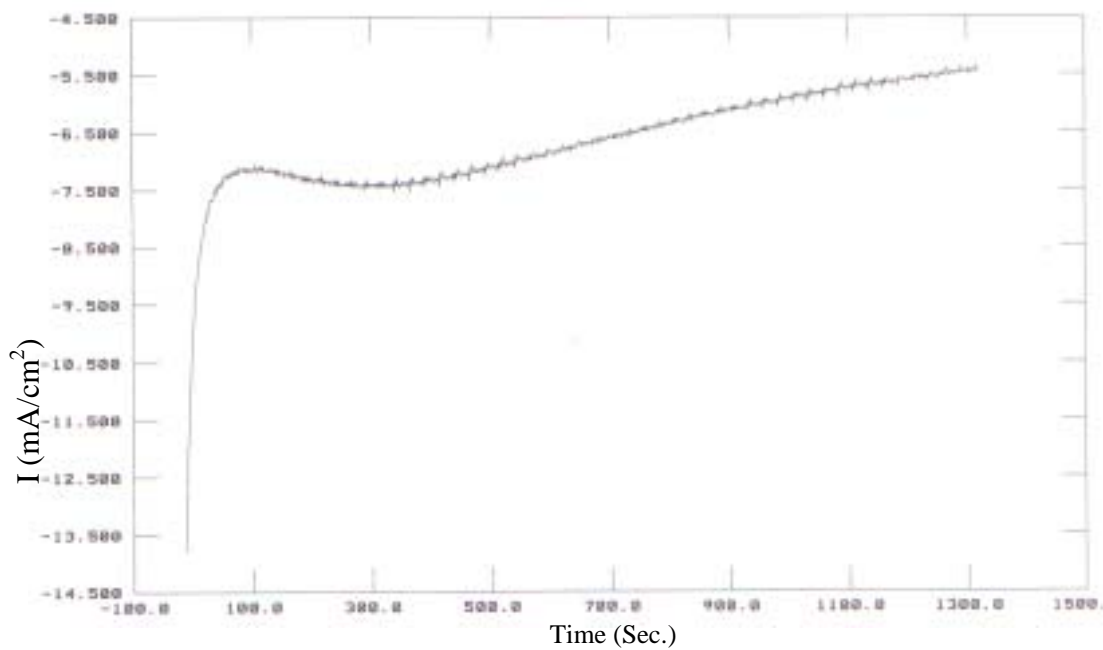

Figure 12. Variation of current density with time (chronoamperometry) during electrochemical oxidation of HgBaCaCuO alloyed film.

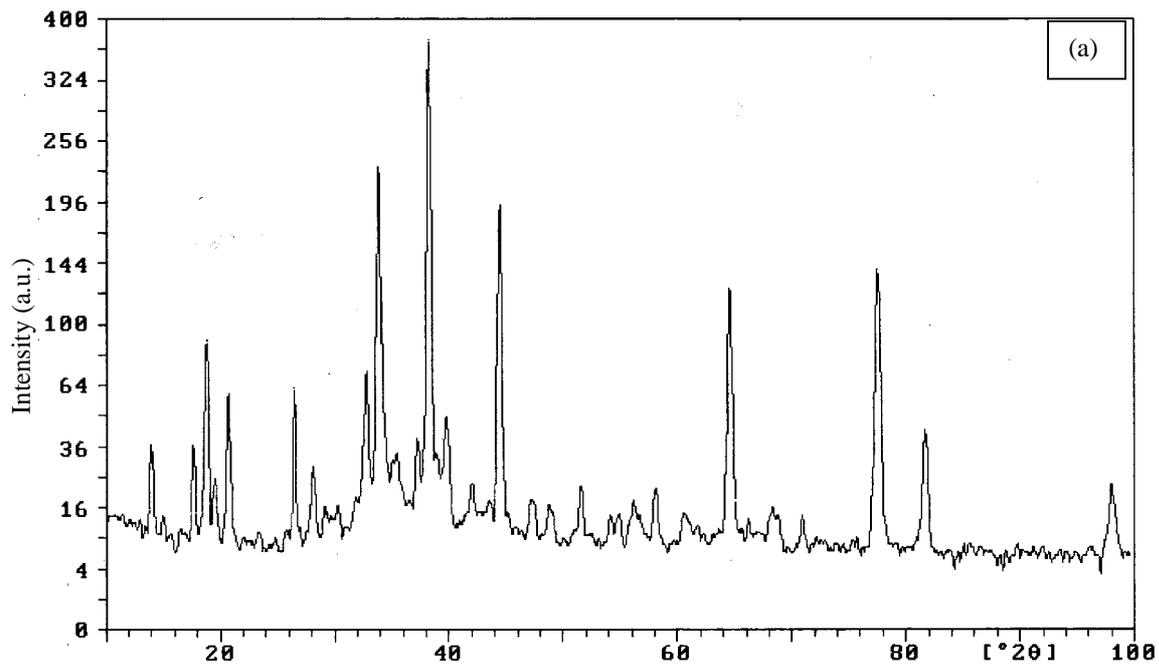

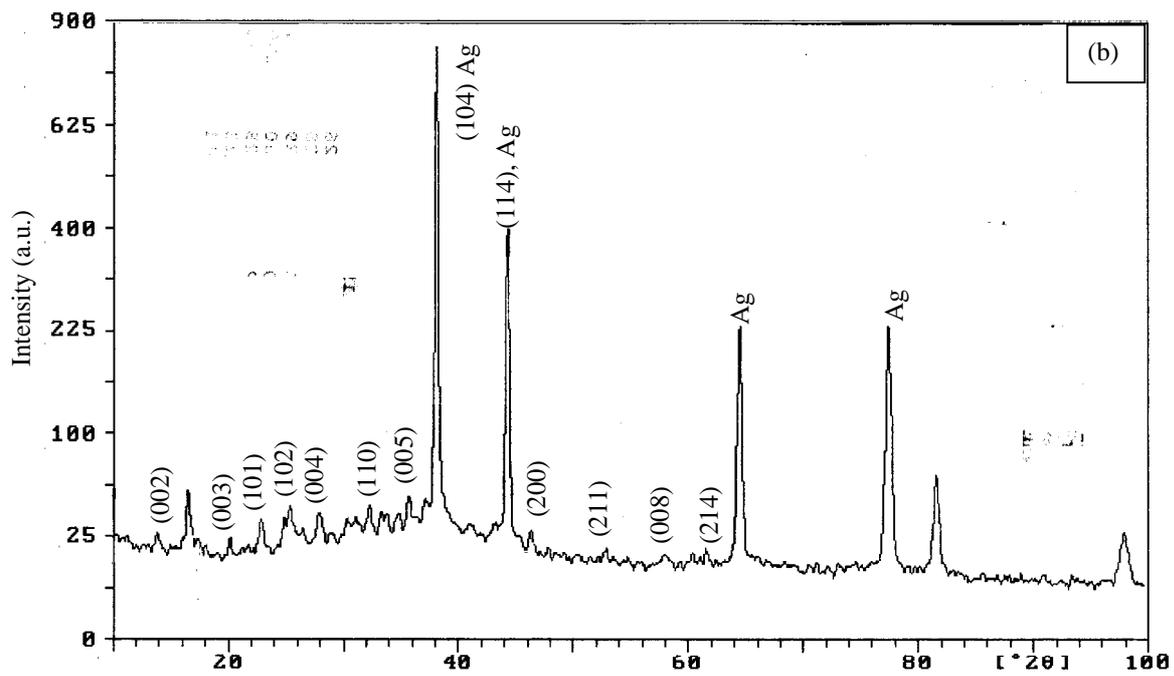

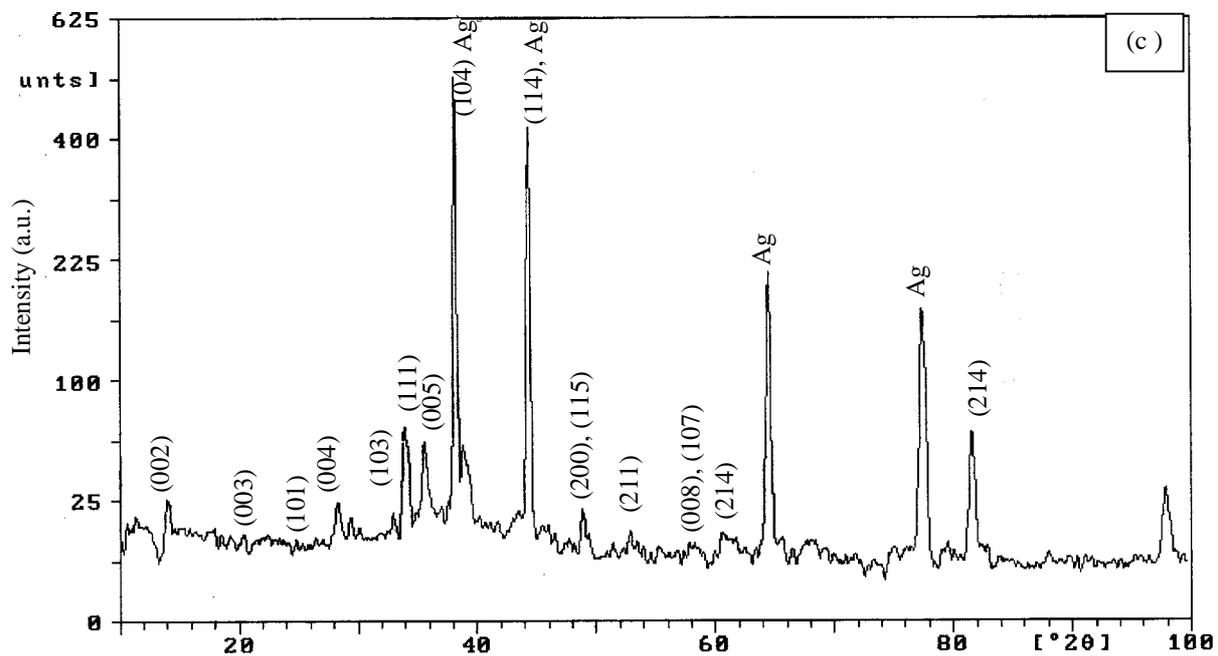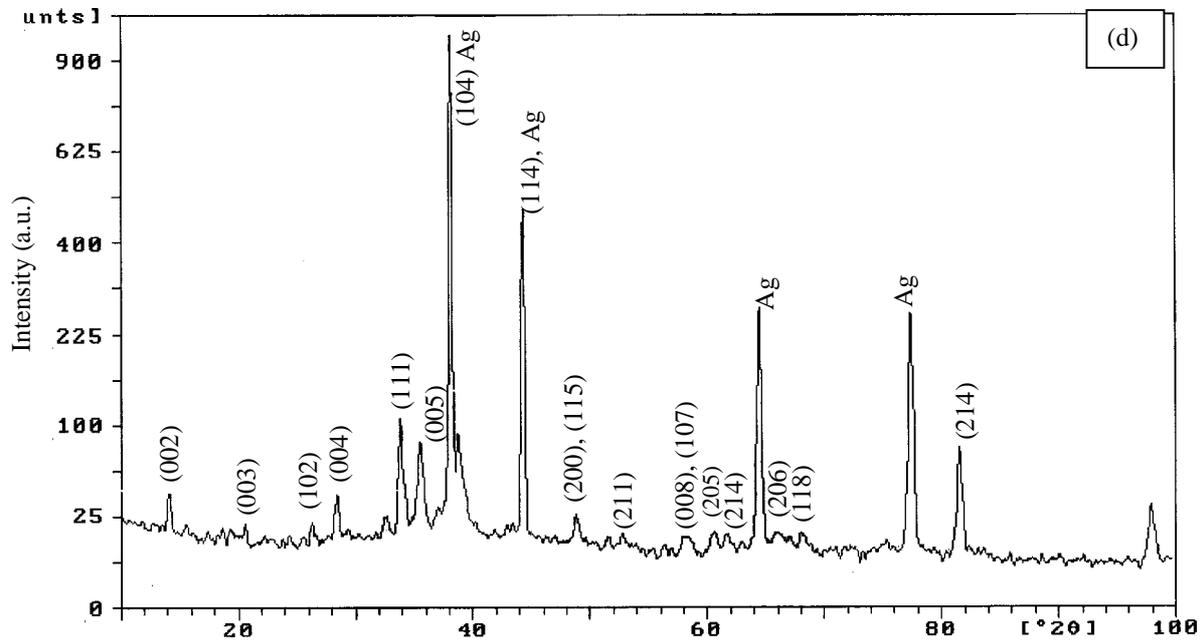

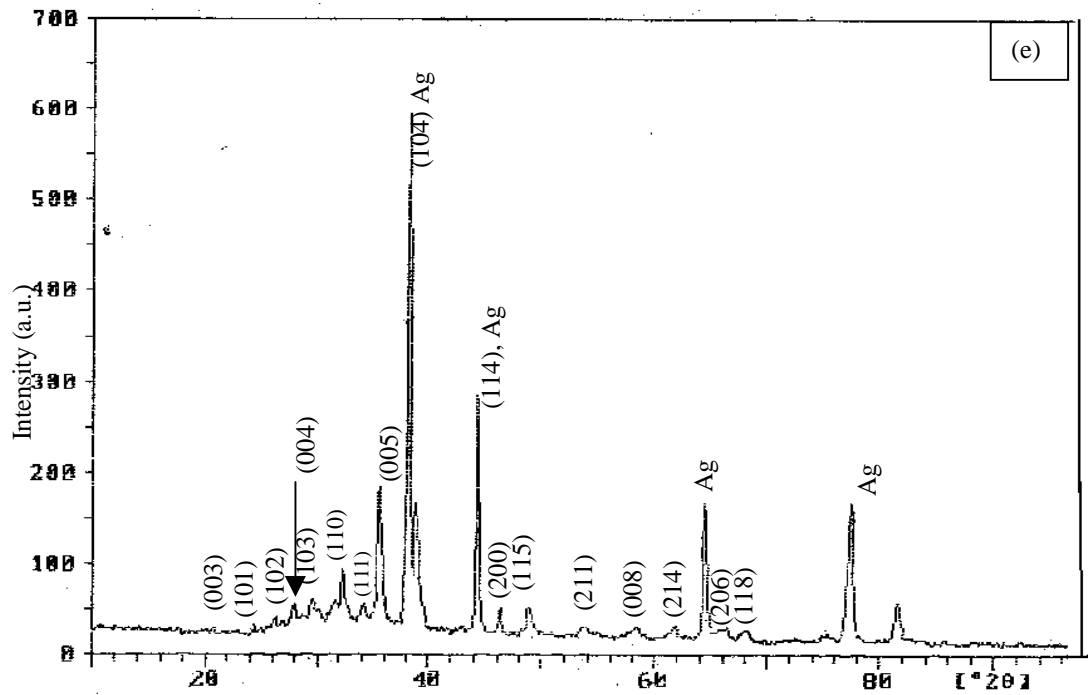

Figure 13 (a-e): The X-ray diffraction pattern for (a) as-deposited HgBaCaCu alloyed film and (b) 20 min.; (c) 24 min.; (d) 28 min.; and (e) 30 minutes electrochemically oxidized films

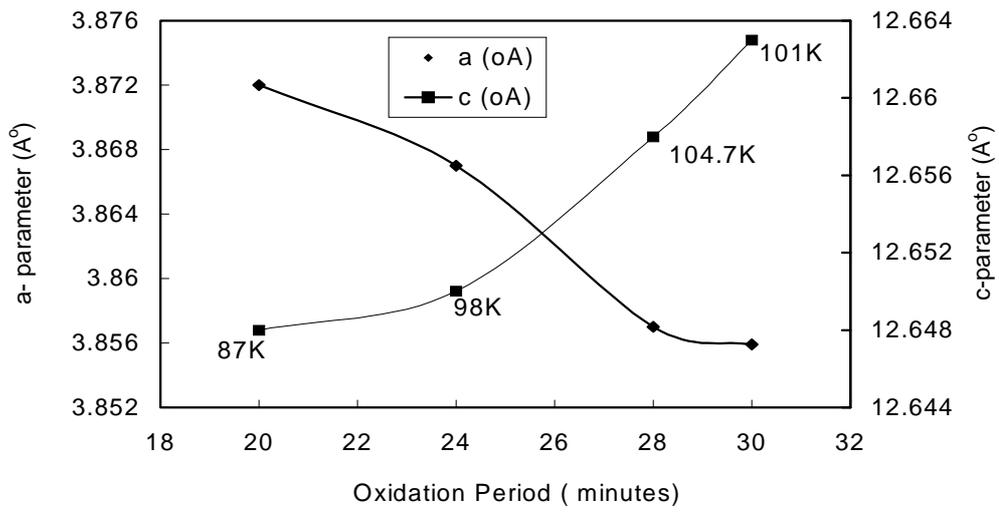

Figure 14 Variation of *a* and *c* parameter with oxidation period

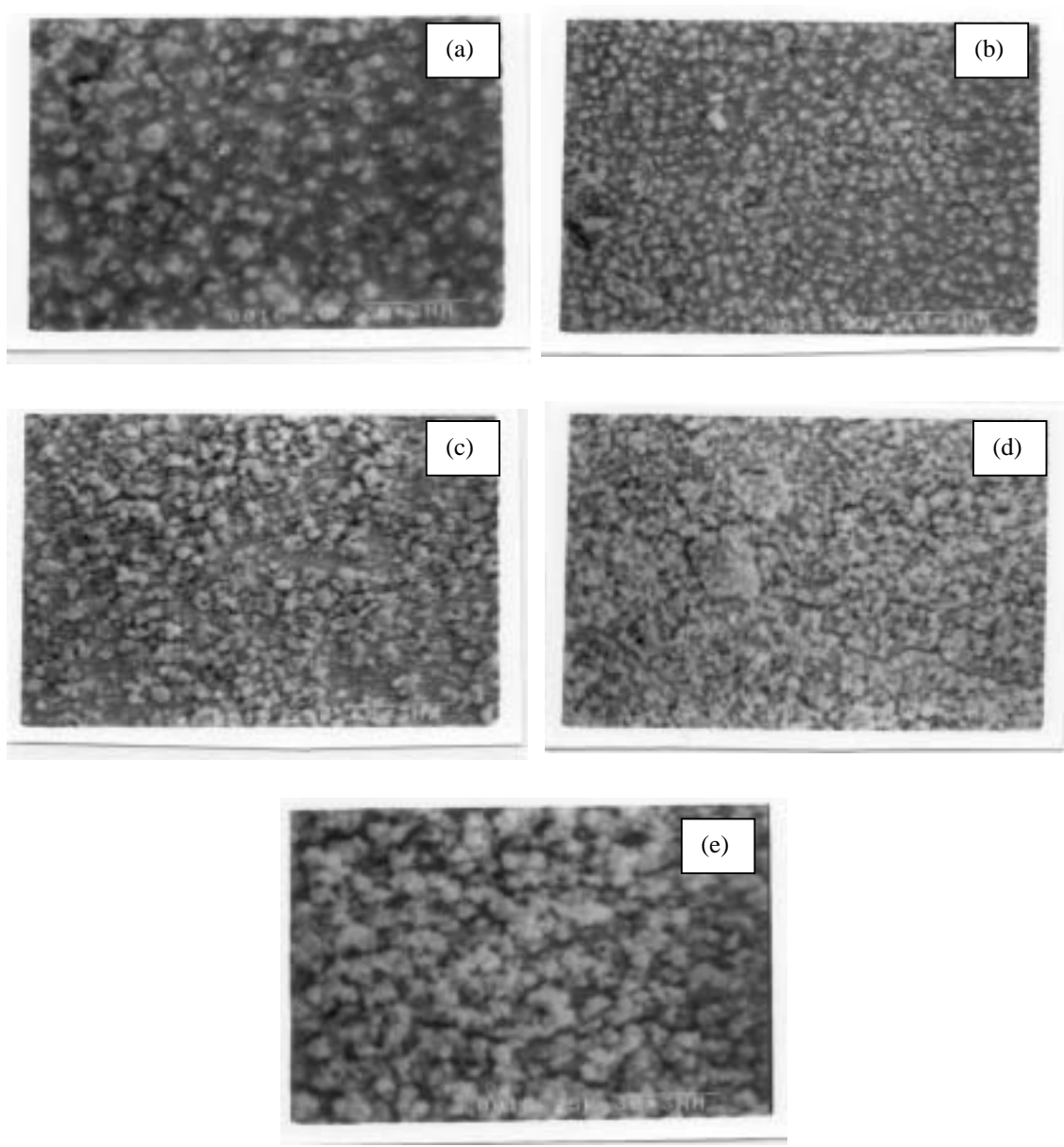

Figure 15 (a-e): Scanning electron micrographs for
(a) as-deposited HgBaCaCu alloyed film and electrochemically oxidized films for (b) 20 min.; (c) 24 min.; (d) 28 min.; and (e) 30 minutes.

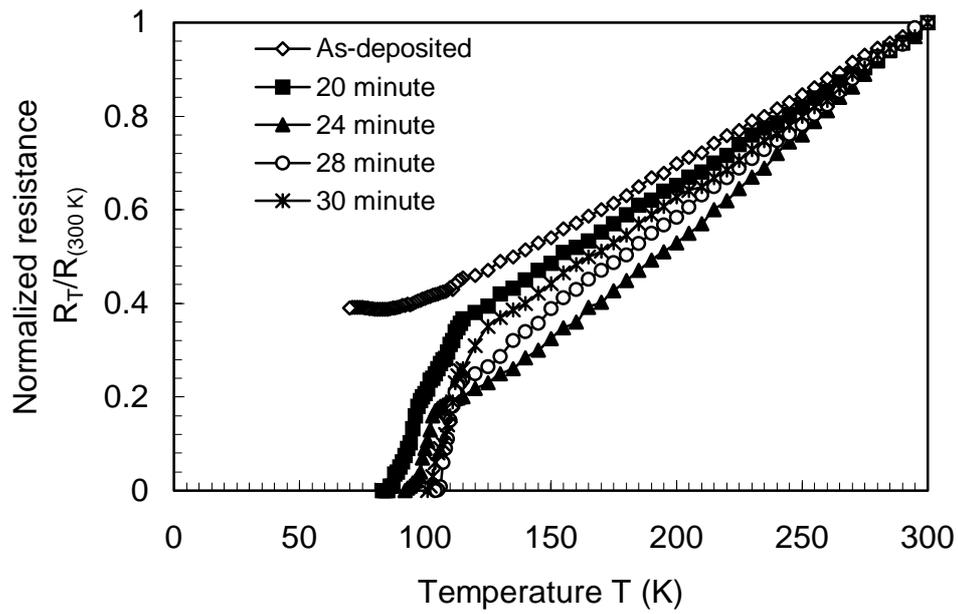

Figure 16. Temperature dependence of normalized resistivity of the Hg-Ba-Ca-CuO films electrochemically oxidized for different periods

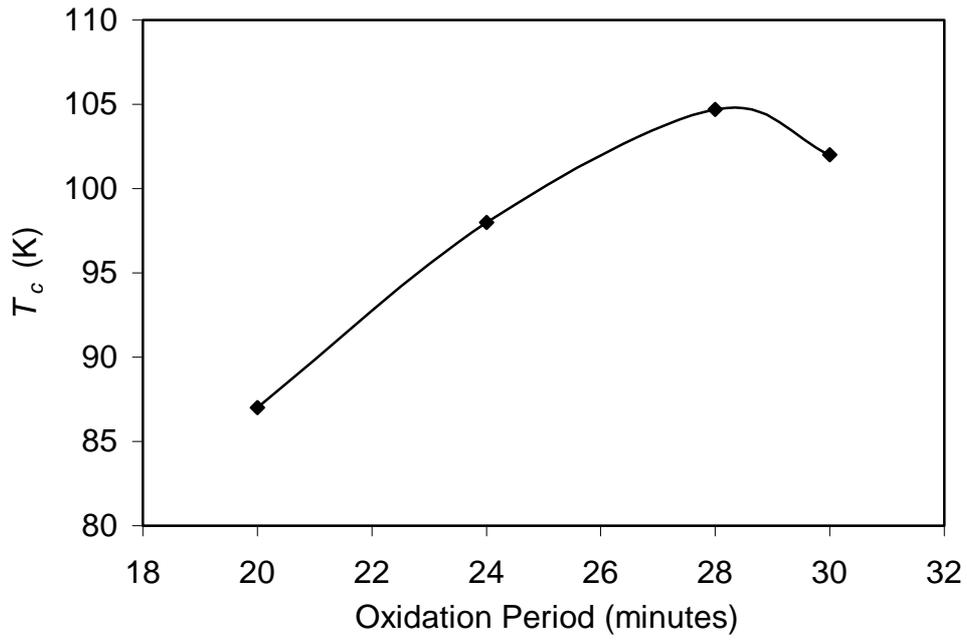

Figure 17(a) Transition temperatures achieved for the films oxidized for different electrochemical oxidation period

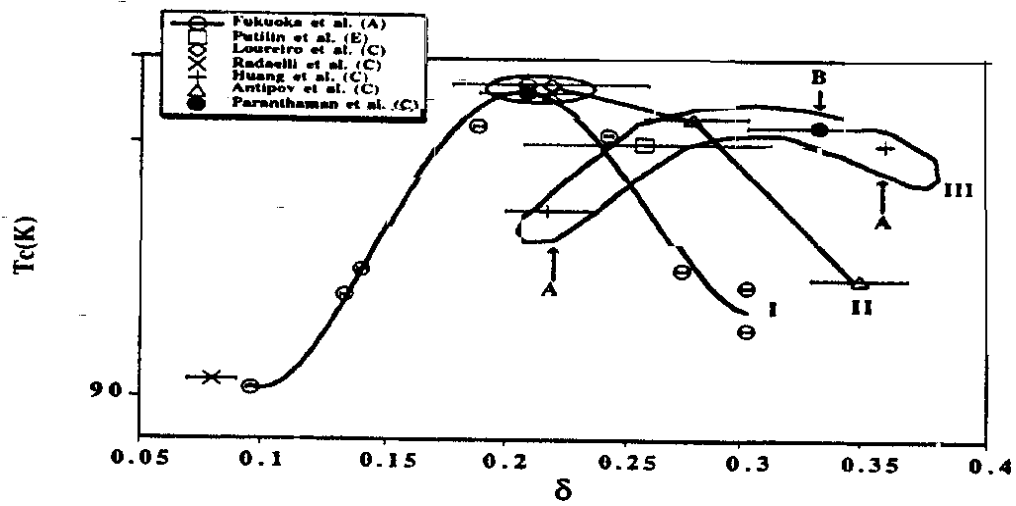

Fig. 17 (b). Variation in $T_c$ with $\delta$ as observed by Fukuoka [55].